\documentclass[12pt]{article}
\pdfoutput=1
\usepackage{amsmath,amssymb,amsthm,amsxtra,overpic,bbm,bm,epsfig,subfigure}
\usepackage{mathrsfs}
\usepackage{graphicx}
\usepackage{color}
\usepackage{comment}
\usepackage{epstopdf}
\usepackage{float}
\usepackage{slashed,stmaryrd}

\def\thefootnote{\fnsymbol{footnote}}

\begin{document}

\vspace{0.2cm}

\begin{center}
{\Large\bf Renormalization-Group Equations of Neutrino Masses and Flavor Mixing Parameters in Matter}
\end{center}

\vspace{0.2cm}

\begin{center}
{\bf Zhi-zhong Xing~$^{a,b,c}$} \footnote{E-mail: xingzz@ihep.ac.cn},
\quad
{\bf Shun Zhou~$^{a,b}$} \footnote{E-mail: zhoush@ihep.ac.cn},
\quad
{\bf Ye-Ling Zhou~$^{d}$} \footnote{E-mail: ye-ling.zhou@durham.ac.uk}
\\
{\small $^a$Institute of High Energy Physics, Chinese Academy of
Sciences, Beijing 100049, China \\
$^b$School of Physical Sciences, University of Chinese Academy of Sciences, Beijing 100049, China \\
$^c$Center for High Energy Physics, Peking University, Beijing 100871, China \\
$^d$Institute for Particle Physics Phenomenology, Department of Physics,
Durham University, \\ Durham DH1 3LE, United Kingdom}
\end{center}

\vspace{1.5cm}

\begin{abstract}
We borrow the general idea of renormalization-group equations (RGEs) to understand
how neutrino masses and flavor mixing parameters evolve when
neutrinos propagate in a medium, highlighting a meaningful possibility that the
genuine flavor quantities in vacuum can be extrapolated from their
matter-corrected counterparts to be measured in some realistic neutrino oscillation experiments. Taking the matter parameter
$a \equiv 2\sqrt{2} \ G^{}_{\rm F} N^{}_e E$ to be an arbitrary scale-like
variable with $N^{}_e$ being the net electron number density and $E$
being the neutrino beam energy, we derive a complete set of differential
equations for the effective neutrino mixing matrix $V$ and the effective
neutrino masses $\widetilde{m}^{}_i$ (for $i = 1, 2, 3$). Given the standard parametrization of $V$, the RGEs for $\{\widetilde{\theta}^{}_{12}, \widetilde{\theta}^{}_{13}, \widetilde{\theta}^{}_{23}, \widetilde{\delta}\}$
in matter are formulated for the first time. We demonstrate some useful
differential invariants which retain the same form from vacuum to matter,
including the well-known Naumov and Toshev relations. The RGEs
of the partial $\mu$-$\tau$ asymmetries, the off-diagonal asymmetries and
the sides of unitarity triangles of $V$ are also obtained as a by-product.
\end{abstract}

\begin{flushleft}
\hspace{0.8cm} PACS number(s): 14.60.Pq, 25.30.Pt
\end{flushleft}

\def\thefootnote{\arabic{footnote}}
\setcounter{footnote}{0}

\newpage

\section{Introduction}

Now it has been firmly established by a number of elegant neutrino oscillation experiments in the past two decades that neutrinos are actually massive and lepton flavors are significantly mixed~\cite{Patrignani:2016xqp}. In the framework of three generations of massive neutrinos, the phenomena of lepton flavor mixing can well be described by the $3\times 3$ unitary matrix $U$, which is conventionally parametrized in terms of three flavor mixing angles $\{\theta^{}_{12}, \theta^{}_{13}, \theta^{}_{23}\}$ and one CP-violating phase $\delta$
\footnote{Throughout this work we do not consider the possible Majorana phases, simply because they are irrelevant to neutrino oscillations in both vacuum and matter.}.
In the standard parametrization of $U$, we have~\cite{Patrignani:2016xqp}
\begin{eqnarray}\label{eq:PMNS}
U = \left(\begin{matrix} U^{}_{e1} & U^{}_{e2} & U^{}_{e3} \\ U^{}_{\mu 1} & U^{}_{\mu 2} & U^{}_{\mu 3} \\ U^{}_{\tau 1} & U^{}_{\tau 2} & U^{}_{\tau 3}  \end{matrix}\right) = \left(\begin{matrix} c^{}_{12} c^{}_{13} & s^{}_{12} c^{}_{13} & s^{}_{13} \\  - c^{}_{12} s^{}_{13} s^{}_{23} - s^{}_{12} c^{}_{23}  e^{-{\rm i}\delta} & - s^{}_{12} s^{}_{13} s^{}_{23} + c^{}_{12} c^{}_{23} e^{-{\rm i}\delta} & c^{}_{13} s^{}_{23} \\ - c^{}_{12} s^{}_{13} c^{}_{23} + s^{}_{12} s^{}_{23} e^{-{\rm i}\delta}
& - s^{}_{12} s^{}_{13} c^{}_{23} - c^{}_{12} s^{}_{23} e^{-{\rm i}\delta}  & c^{}_{13} c^{}_{23}\end{matrix}\right) \; ,
\end{eqnarray}
where $c^{}_{ij} \equiv \cos\theta^{}_{ij}$ and $s^{}_{ij} \equiv \sin\theta^{}_{ij}$ (for $ij = 12, 13, 23$) have been introduced. The latest global-fit analysis of neutrino oscillation data yield~\cite{Esteban:2016qun} the best-fit values of four flavor mixing parameters $\theta^{}_{12} \approx 33.6^\circ$, $\theta^{}_{13} \approx 8.5^\circ$, $\theta^{}_{23} \approx 47.2^\circ$ and $\delta \approx 234^\circ$, and those of two independent neutrino mass-squared differences $\Delta^{}_{21} \equiv m^2_2 - m^2_1 \approx 7.40\times 10^{-5}~{\rm eV}^2$ and $\Delta^{}_{31} \equiv m^2_3 - m^2_1 \approx  2.49\times 10^{-3}~{\rm eV}^2$. Although there exists currently a slight preference for the normal neutrino mass ordering (NO, i.e., $\Delta^{}_{31} > 0$), the inverted mass ordering (IO, i.e., $\Delta^{}_{31} < 0$) is still allowed. Some preliminary hints on the maximal CP-violating phase $\delta \approx 270^\circ$ arise from the long-baseline accelerator neutrino experiments~\cite{Abe:2017uxa, Abe:2017vif}, which needs to be confirmed when more data are available in the near future.

The determination of neutrino mass ordering and leptonic CP-violating phase $\delta$ in the long-baseline accelerator neutrino oscillation experiments calls for an excellent understanding of the Mikheyev-Smirnov-Wolfenstein (MSW) matter effects~\cite{Wolfenstein:1977ue, Mikheev:1986gs, Mikheev:1986wj}, which becomes crucially important when the neutrino beam propagates in the Earth matter for a long distance. For the three-flavor neutrino oscillations in matter, the effective Hamiltonian reads
\begin{eqnarray} \label{eq:Hm}
H^{}_{\rm m} = \frac{1}{2E} \left[ U \left(\begin{matrix} m^2_1 & 0 & 0 \\ 0 & m^2_2 & 0 \\ 0 & 0 & m^2_3 \end{matrix}\right) U^\dagger + \left(\begin{matrix} a & 0 & 0 \\ 0 & 0 & 0 \\ 0 & 0 & 0 \end{matrix}\right) \right] \equiv \frac{1}{2E} V \left(\begin{matrix} \widetilde{m}^2_1 & 0 & 0 \\ 0 & \widetilde{m}^2_2 & 0 \\ 0 & 0 & \widetilde{m}^2_3 \end{matrix}\right) V^\dagger \; ,
\end{eqnarray}
where $a \equiv 2\sqrt{2} \ G^{}_{\rm F} N^{}_e E$ with $G^{}_{\rm F} = 1.166\times 10^{-5}~{\rm GeV}^{-2}$ being the Fermi constant, $N^{}_e$ the net electron number density and $E$ the neutrino beam energy. For antineutrino oscillations in matter, one may simply replace $U$ by $U^*$ and $a$ by $-a$ in the effective Hamiltonian. In Eq.~(\ref{eq:Hm}) the effective flavor mixing matrix $V$ and neutrino masses $\widetilde{m}^{}_i$ (for $i = 1, 2, 3$) in matter have been defined. For any realistic profile of the matter density, it is possible to numerically calculate neutrino oscillation probabilities by solving the evolution equations of neutrino flavor states. However, the analytical relations or identities between the effective mixing parameters in matter and the fundamental ones in vacuum are very helpful. For instance, the well-known Naumov~\cite{Naumov:1991ju, Harrison:1999df, Xing:2000gg, Xing:2001bg} and Toshev~\cite{Toshev:1991ku} relations can be summarized as~\cite{Xing:2016ymg}
\begin{eqnarray}\label{eq:NS}
\frac{\widetilde{\cal J}}{{\cal J}} = \left|\frac{V^{}_{e1}}{U^{}_{e1}}\right| \left|\frac{V^{}_{e2}}{U^{}_{e2}}\right| \left|\frac{V^{}_{e3}}{U^{}_{e3}}\right| = \frac{\Delta^{}_{12} \Delta^{}_{23} \Delta^{}_{31}}{\widetilde{\Delta}^{}_{12} \widetilde{\Delta}^{}_{23} \widetilde{\Delta}^{}_{31}} \; ,
\end{eqnarray}
where $\widetilde{\Delta}^{}_{ij} \equiv \widetilde{m}^2_i - \widetilde{m}^2_j$ (for $ij = 12, 23, 31$), and the Jarlskog invariant in vacuum \cite{Jarlskog:1985ht}
and its counterpart in matter are defined via
\begin{eqnarray}\label{eq:Jarl}
{\rm Im}\left(U^{}_{\alpha i} U^{}_{\beta j} U^*_{\alpha j} U^*_{\beta i}\right)
& = & {\cal J} \sum_\gamma \sum_k \epsilon^{}_{\alpha \beta \gamma} \epsilon^{}_{ijk}
\; , \nonumber \\
{\rm Im}\left(V^{}_{\alpha i} V^{}_{\beta j} V^*_{\alpha j} V^*_{\beta i}\right)
& = & \widetilde{\cal J} \sum_\gamma \sum_k \epsilon^{}_{\alpha \beta \gamma} \epsilon^{}_{ijk} \; ,
\end{eqnarray}
with $\epsilon^{}_{\alpha \beta \gamma}$ and $\epsilon^{}_{ijk}$ being totally antisymmetric tensors, and $(\alpha, \beta, \gamma)$ and $(i, j, k)$ being cyclic permutations of $(e, \mu, \tau)$ and $(1, 2, 3)$, respectively. Moreover, some interesting sum rules for $\widetilde{m}^{2}_i$ and the matrix elements of $V$ have been derived in Ref.~\cite{Xing:2001bg} and used to study the unitarity triangles of $V$ in matter~\cite{Zhang:2004hf, Xing:2005gk, Xing:2015wzz}.

In this paper we emphasize that the dependence of the effective mixing parameters $V^{}_{\alpha i}$ and $\widetilde{m}^2_i$ (for $\alpha = e, \mu, \tau$ and $i = 1, 2, 3$) on the matter term $a$ can perfectly be described by a complete set of differential equations, which are analogous to the renormalization-group equations (RGEs) associated with the dependence of fundamental parameters on the renormalization energy scale or distance in quantum field theories \cite{Stueckelberg, GellMann}, solid-state physics \cite{Wilson1, Wilson2} and other fields of modern physics \cite{Others}
\footnote{Although $E$ in $a$ denotes the kinetic energy of a neutrino beam,
it is also a reflection of the energy scale associated with weak charged-current interactions between the electron neutrino (or antineutrino) flavor and the
electrons in matter. In this sense it should be reasonable to treat $a$ as a scale-like variable.}.
Although this interesting analogy has already been pointed out in Refs.~\cite{Chiu:2010da, Chiu:2017ckv}, it deserves some highlights and a further study. We argue that the introduction of effective neutrino mass-squared differences and effective flavor mixing parameters guarantees the form invariance of neutrino oscillation probabilities in vacuum and in a medium with arbitrary values of $a$. Such a form invariance (or self-similarity) exactly reflects the spirit of the RGEs
\cite{Stueckelberg, GellMann, Wilson1}, and thus it implies the validity of the RGE-like approach
for neutrino oscillations in matter.

It is worth remarking that our present work differs from Refs.~\cite{Chiu:2010da, Chiu:2017ckv} in several nontrivial aspects. First, we explain why the RGE language can be applied to the description of neutrino oscillation parameters in matter changing with the scale-like variable $a$. With this key point in mind, we derive the RGEs for neutrino masses $\widetilde{m}^{}_i$, the squared-moduli of flavor mixing matrix elements $|V^{}_{\alpha i}|^2$ and even the matrix elements $V^{}_{\alpha i}$ themselves. Second, we demonstrate that the standard parametrization of $V$ is most convenient for the derivation of the RGEs of three flavor mixing angles and one CP-violating phase, because it makes the first row of $V$ so simple that the coherent forward scattering between electrons and electron
neutrinos (or antineutrinos) via weak charged-current interactions can be described in a very simple way. The RGEs of such mixing parameters will also be numerically solved, and the salient features of their evolution with respect to the matter parameter $a$ will be discussed. Third, the RGEs of $\widetilde{\cal J}$ and some other interesting quantities, such as the partial $\mu$-$\tau$ asymmetries, the off-diagonal asymmetries and the sides of unitarity triangles of $V$, are derived as a by-product. Fourth, we compare the newly obtained differential results with some previously
obtained integral results, and highlight the complementarity of both approaches in describing and understanding matter effects on neutrino oscillations.

In particular, we highlight that the RGEs for neutrinos running in matter may hopefully provide
a meaningful possibility that the genuine (or fundamental) flavor quantities in vacuum can be extrapolated from their matter-corrected (or effective) counterparts to be measured in some realistic neutrino oscillation experiments.

The remaining part of our paper is structured as follows. In Section~2, we derive the RGEs of the effective mixing parameters and neutrino masses explicitly and establish our conventions and notations. Adopting the standard parametrization of $V$, we further present the explicit expressions of the RGEs for the mixing parameters $\{\widetilde{\theta}^{}_{12}, \widetilde{\theta}^{}_{13}, \widetilde{\theta}^{}_{23}, \widetilde{\delta}\}$ in Section~3. Section~4 is devoted to further discussions on the RGEs of other phenomenologically interesting quantities. Finally, we summarize our main results in Section~5.

\section{Renormalization-Group Equations}

The essential idea of ours is to study the dependence of the flavor mixing parameters on the
scale-like matter term $a$ by following the normal RGE approach. Differentiating both sides of Eq.~(\ref{eq:Hm}) with respect to $a$, we immediately obtain
\begin{eqnarray}\label{eq:RGE}
\dot{D} + \left[V^\dagger \dot{V}, D\right] = V^\dagger \left(\begin{matrix} 1 & 0 & 0 \\ 0 & 0 & 0 \\ 0 & 0 & 0 \end{matrix}\right) V = \left(\begin{matrix} |V^{}_{e 1}|^2 & V^*_{e 1} V^{}_{e 2} & V^*_{e 1} V^{}_{e 3} \\ V^*_{e 2} V^{}_{e 1} & |V^{}_{e 2}|^2 & V^*_{e 2} V^{}_{e 3} \\ V^*_{e 3} V^{}_{e 1} & V^*_{e 3} V^{}_{e 2} & |V^{}_{e 3}|^2 \end{matrix}\right) \; ,
\end{eqnarray}
where the derivatives are denoted by overhead dots, $D \equiv {\rm diag}\{\widetilde{m}^2_1, \widetilde{m}^2_2, \widetilde{m}^2_3\}$ and $[A, B] \equiv AB - BA$ is the commutator of two matrices $A$ and $B$. Since the diagonal matrix elements of the commutator are always vanishing, it is straightforward to get
\begin{eqnarray}\label{eq:diag}
\frac{{\rm d} \widetilde{m}^2_i}{{\rm d}a} = |V^{}_{e i}|^2 \; ,
\end{eqnarray}
for $i = 1, 2, 3$ by equating the diagonal elements on both sides of Eq.~(\ref{eq:RGE}); and
\begin{eqnarray}
\label{eq:offdiag}
\sum_{\alpha} V^*_{\alpha i} \dot{V}^{}_{\alpha j} &=& V^*_{e i} V^{}_{e j} \widetilde{\Delta}^{-1}_{ji}  \; ,
\end{eqnarray}
for $i \neq j$ by identifying the off-diagonal elements. In addition, we have a few useful identities from the normalization and orthogonality conditions for the unitary matrix $V$, namely,
\begin{eqnarray}\label{eq:norm}
\sum_{i} \left(V^*_{\alpha i} \dot{V}^{}_{\alpha i} + \dot{V}^*_{\alpha i} V^{}_{\alpha i}\right) = \sum_{\alpha} \left(V^*_{\alpha i} \dot{V}^{}_{\alpha i} + \dot{V}^*_{\alpha i} V^{}_{\alpha i}\right) = 0 \; ,
\end{eqnarray}
which can be recast into $\displaystyle\sum_i {\rm Re}\left(V^*_{\alpha i} \dot{V}^{}_{\alpha i} \right) = \sum_\alpha {\rm Re} \left(V^*_{\alpha i} \dot{V}^{}_{\alpha i} \right) = 0$, where $i = 1, 2, 3$ and $\alpha = e, \mu, \tau$ are implied; and
\begin{eqnarray}\label{eq:orth}
\sum_{i} \left(V^*_{\alpha i} \dot{V}^{}_{\beta i} + \dot{V}^*_{\alpha i} V^{}_{\beta i}\right) = \sum_{\alpha} \left(V^*_{\alpha i} \dot{V}^{}_{\alpha j} + \dot{V}^*_{\alpha i} V^{}_{\alpha j}\right) = 0 \; ,
\end{eqnarray}
where $\alpha \neq \beta$ and $i \neq j$ should be noticed in the first and second identities, respectively. With the help of the above equations, we are now ready to derive the RGEs for the matrix elements of $V$ and the relevant rephasing invariants. The main results are summarized below:
\begin{itemize}
\item Starting with the orthogonality condition $V^*_{\alpha 1} V^{}_{\beta 1} + V^*_{\alpha 2} V^{}_{\beta 2} + V^*_{\alpha 3} V^{}_{\beta 3} = 0$, or equivalently,
\begin{eqnarray}\label{eq:ortheqv}
\sum_{j \neq i} V^*_{\alpha j} V^{}_{\beta j} = \delta^{}_{\alpha \beta} - V^*_{\alpha i} V^{}_{\beta i} \; ,
\end{eqnarray}
we multiply both sides of Eq.~(\ref{eq:ortheqv}) by $\dot{V}^{}_{\alpha i}$ and sum over the flavor index $\alpha$. Then, by using Eq.~(\ref{eq:offdiag}), we arrive at
\begin{eqnarray}\label{eq:elements}
\dot{V}^{}_{\beta i} = \sum_\alpha \dot{V}^{}_{\alpha i} V^*_{\alpha i} V^{}_{\beta i} + \sum_{j\neq i} V^{}_{ei} V^*_{e j} V^{}_{\beta j} \widetilde{\Delta}^{-1}_{ij} \; .
\end{eqnarray}
Note that the first term on the right-hand side of Eq.~(\ref{eq:elements}) is rephasing-dependent, and it can be arranged to vanish in a special phase convention without altering any physical results~\cite{Chiu:2017ckv}, as one has noticed in deriving the RGEs of quark flavor mixing parameters~\cite{Kielanowski:2008wm}. We shall confirm that the terms associated with $\displaystyle\sum_\alpha \dot{V}^{}_{\alpha i} V^*_{\alpha i}$ can always be cancelled out in our subsequent calculations.

\item Since there will be unphysical phases in the mixing matrix $V$, it is more interesting to present the RGEs for the rephasing invariants. The simplest ones are just the squared-moduli $|V^{}_{\alpha i}|^2$, whose RGEs can be directly derived from Eq.~(\ref{eq:elements}):
\begin{eqnarray}\label{eq:Vai2}
\frac{{\rm d}}{{\rm d}a} |V^{}_{\alpha i}|^2 = \left( \frac{{\rm d}}{{\rm d}a}V^*_{\alpha i}\right) V^{}_{\alpha i} + V^*_{\alpha i} \left(\frac{{\rm d}}{{\rm d}a} V^{}_{\alpha i} \right) = 2 \sum_{j\neq i} {\rm Re}\left[V^{}_{e i} V^{}_{\alpha j} V^*_{e j} V^*_{\alpha i}\right] \widetilde{\Delta}^{-1}_{ij} \; ,
\end{eqnarray}
where the second identity in Eq.~(\ref{eq:norm}) has been used. In principle, the RGEs in Eqs.~(\ref{eq:diag}) and (\ref{eq:Vai2}) are sufficient to investigate the evolution of all physical quantities with respect to the matter term $a$, since the moduli $|V^{}_{\alpha i}|$ of four independent matrix elements can unambiguously determine all three mixing angles and one CP-violating phase. Specifying $\alpha = e$ and $i = 1, 2, 3$ in Eq.~(\ref{eq:Vai2}), we explicitly have
\begin{eqnarray}\label{eq:Vei2}
\frac{{\rm d}}{{\rm d}a} |V^{}_{e1}|^2 &=& 2 |V^{}_{e1}|^2 \left( |V^{}_{e2}|^2 \widetilde{\Delta}^{-1}_{12} - |V^{}_{e3}|^2 \widetilde{\Delta}^{-1}_{31} \right) \; , \nonumber \\
\frac{{\rm d}}{{\rm d}a} |V^{}_{e2}|^2 &=& 2 |V^{}_{e2}|^2 \left(|V^{}_{e3}|^2 \widetilde{\Delta}^{-1}_{23} - |V^{}_{e1}|^2 \widetilde{\Delta}^{-1}_{12} \right) \; , \nonumber \\
\frac{{\rm d}}{{\rm d}a} |V^{}_{e3}|^2 &=& 2 |V^{}_{e3}|^2 \left( |V^{}_{e1}|^2 \widetilde{\Delta}^{-1}_{31} - |V^{}_{e2}|^2 \widetilde{\Delta}^{-1}_{23} \right) \; ,
\end{eqnarray}
together with
\begin{eqnarray}\label{eq:dDeltaij}
\frac{{\rm d}}{{\rm d}a} \widetilde{\Delta}^{}_{12} &=& |V^{}_{e1}|^2 - |V^{}_{e2}|^2 \; , \nonumber \\
\frac{{\rm d}}{{\rm d}a} \widetilde{\Delta}^{}_{23} &=& |V^{}_{e2}|^2 - |V^{}_{e3}|^2 \; , \nonumber \\
\frac{{\rm d}}{{\rm d}a} \widetilde{\Delta}^{}_{31} &=& |V^{}_{e3}|^2 - |V^{}_{e1}|^2 \; ,
\end{eqnarray}
from Eq.~(\ref{eq:diag}). Note that the RGEs in Eqs.~(\ref{eq:Vei2}) and (\ref{eq:dDeltaij}) are closed for $\{|V^{}_{e1}|^2, |V^{}_{e2}|^2, |V^{}_{e3}|^2\}$ and $\{\widetilde{\Delta}^{}_{12}, \widetilde{\Delta}^{}_{23}, \widetilde{\Delta}^{}_{31}\}$, and completely symmetric under the cyclic permutations among the subscripts $(1, 2, 3)$. Due to the normalization condition $|V^{}_{e1}|^2 + |V^{}_{e2}|^2 + |V^{}_{e3}|^2 = 1$ and the identity $\widetilde{\Delta}^{}_{12} + \widetilde{\Delta}^{}_{23} + \widetilde{\Delta}^{}_{31} =0$, there are only four independent differential equations in Eqs.~(\ref{eq:Vei2}) and (\ref{eq:dDeltaij}). However, two redundant equations have been included in order to put them in a more symmetric form. For comparison, we quote the existing sum rules for $|V^{}_{e i}|^2$ and $|U^{}_{e i}|^2$ (for $i = 1, 2, 3$) from Ref.~\cite{Xing:2005gk}:
\begin{eqnarray}\label{eq:sumrule}
|V^{}_{e 1}|^2 &=& \frac{\widehat{\Delta}^{}_{21} \widehat{\Delta}^{}_{31}}{\widetilde{\Delta}^{}_{21} \widetilde{\Delta}^{}_{31}} |U^{}_{e1}|^2 + \frac{\widehat{\Delta}^{}_{11} \widehat{\Delta}^{}_{31}}{\widetilde{\Delta}^{}_{21} \widetilde{\Delta}^{}_{31}} |U^{}_{e2}|^2 + \frac{\widehat{\Delta}^{}_{11} \widehat{\Delta}^{}_{21}}{\widetilde{\Delta}^{}_{21} \widetilde{\Delta}^{}_{31}} |U^{}_{e3}|^2 \; , \nonumber \\
|V^{}_{e 2}|^2 &=& \frac{\widehat{\Delta}^{}_{22} \widehat{\Delta}^{}_{32}}{\widetilde{\Delta}^{}_{12} \widetilde{\Delta}^{}_{32}} |U^{}_{e1}|^2 + \frac{\widehat{\Delta}^{}_{12} \widehat{\Delta}^{}_{32}}{\widetilde{\Delta}^{}_{12} \widetilde{\Delta}^{}_{32}} |U^{}_{e2}|^2 + \frac{\widehat{\Delta}^{}_{12} \widehat{\Delta}^{}_{22}}{\widetilde{\Delta}^{}_{12} \widetilde{\Delta}^{}_{32}} |U^{}_{e3}|^2 \; , \nonumber \\
|V^{}_{e 3}|^2 &=& \frac{\widehat{\Delta}^{}_{23} \widehat{\Delta}^{}_{33}}{\widetilde{\Delta}^{}_{13} \widetilde{\Delta}^{}_{23}} |U^{}_{e1}|^2 + \frac{\widehat{\Delta}^{}_{13} \widehat{\Delta}^{}_{33}}{\widetilde{\Delta}^{}_{13} \widetilde{\Delta}^{}_{23}} |U^{}_{e2}|^2 + \frac{\widehat{\Delta}^{}_{13} \widehat{\Delta}^{}_{23}}{\widetilde{\Delta}^{}_{13} \widetilde{\Delta}^{}_{23}} |U^{}_{e3}|^2 \; ,
\end{eqnarray}
where $\widehat{\Delta}^{}_{ij} \equiv m^2_i - \widetilde{m}^2_j$. Note that Eq.~(\ref{eq:sumrule}) can be regarded as the formal (integral) solutions to the RGEs of $|V^{}_{e i}|^2$ in Eq.~(\ref{eq:Vei2}) with the mixing matrix elements $|U^{}_{ei}|^2$ and neutrino masses $m^2_i$ in vacuum as initial conditions. Substituting $|V^{}_{e i}|^2$ in Eq.~(\ref{eq:sumrule}) into Eq.~(\ref{eq:dDeltaij}), one can in principle obtain the solutions for $\widetilde{\Delta}^{}_{ij}$.

Given Eqs.~(\ref{eq:Vei2}) and (\ref{eq:dDeltaij}), it is also straightforward to prove~\cite{Chiu:2017ckv}
\begin{eqnarray}\label{eq:Tosh}
\frac{{\rm d}}{{\rm d}a} \left[\ln \left(|V^{}_{e 1}|^2 |V^{}_{e 2}|^2 |V^{}_{e 3}|^2 \widetilde{\Delta}^2_{12} \widetilde{\Delta}^2_{23} \widetilde{\Delta}^2_{31} \right)\right] &=& \sum^3_{i=1} \frac{\rm d}{{\rm d}a} \left(\ln |V^{}_{ei}|^2\right) + \sum_{j>k} \frac{\rm d}{{\rm d}a} \left(\ln \widetilde{\Delta}^2_{jk} \right) = 0 \; , ~~~
\end{eqnarray}
which reproduces the second identity in Eq.~(\ref{eq:NS}). In fact, Eq.~(\ref{eq:Tosh}) indicates that the product $|V^{}_{e1}| |V^{}_{e2}| |V^{}_{e3}| \widetilde{\Delta}^{}_{12} \widetilde{\Delta}^{}_{23} \widetilde{\Delta}^{}_{31}$ is a differential invariant, so its value in matter and that in vacuum (i.e., corresponding to $a = 0$) should be equal to each other. This identity has previously been proved in Ref.~\cite{Kimura:2002wd} by using a different approach.

\item Then we come to the Jarlskog invariant $\widetilde{\cal J}$, whose RGE can be found by starting with its original definition in Eq.~(\ref{eq:Jarl}) and implementing the derivatives of the mixing matrix elements in Eq.~(\ref{eq:elements}). For instance, we have $\widetilde{\cal J} = {\rm Im}\left[V^{}_{e1} V^{}_{\mu 2} V^*_{e2} V^*_{\mu 1}\right]$ and thus its derivative
\begin{eqnarray}\label{eq:dJarl}
\frac{{\rm d}}{{\rm d}a} \widetilde{\cal J} &=& + {\rm Im} \left[\dot{V}^{}_{e1} V^{}_{\mu 2} V^*_{e2} V^*_{\mu 1}\right] + {\rm Im} \left[V^{}_{e1} V^{}_{\mu 2} V^*_{e2} \dot{V}^*_{\mu 1}\right] \nonumber \\
&~& + {\rm Im} \left[V^{}_{e1} \dot{V}^{}_{\mu 2} V^*_{e2} V^*_{\mu 1}\right] + {\rm Im} \left[V^{}_{e1} V^{}_{\mu 2} \dot{V}^*_{e2} V^*_{\mu 1}\right] \; .
\end{eqnarray}
According to Eq.~(\ref{eq:elements}) and its complex conjugate, we can get
\begin{eqnarray}\label{eq:dVei}
\dot{V}^{}_{e1} &=& |V^{}_{e2}|^2 V^{}_{e1} \widetilde{\Delta}^{-1}_{12} - |V^{}_{e3}|^2 V^{}_{e1} \widetilde{\Delta}^{-1}_{31} + \sum_\alpha \dot{V}^{}_{\alpha i} V^*_{\alpha i} V^{}_{e1} \; , \nonumber \\
\dot{V}^*_{e2} &=& |V^{}_{e3}|^2 V^*_{e2} \widetilde{\Delta}^{-1}_{23} - |V^{}_{e1}|^2 V^*_{e2} \widetilde{\Delta}^{-1}_{12} + \sum_\alpha \dot{V}^{}_{\alpha i} V^*_{\alpha i}  V^*_{e2} \; , \nonumber \\
\dot{V}^*_{\mu 1} &=& V^*_{\mu 2} V^*_{e1} V^{}_{e2} \widetilde{\Delta}^{-1}_{12} - V^*_{\mu 3} V^*_{e1} V^{}_{e3} \widetilde{\Delta}^{-1}_{31} + \sum_\alpha \dot{V}^{}_{\alpha i} V^*_{\alpha i}  V^*_{\mu 1} \; , \nonumber \\
\dot{V}^{}_{\mu 2} &=& V^{}_{\mu 3} V^{}_{e2} V^*_{e3} \widetilde{\Delta}^{-1}_{23} - V^{}_{\mu 1} V^{}_{e2} V^*_{e1} \widetilde{\Delta}^{-1}_{12} + \sum_\alpha \dot{V}^{}_{\alpha i} V^*_{\alpha i} V^{}_{\mu 2} \; .
\end{eqnarray}
After inserting Eq.~(\ref{eq:dVei}) into Eq.~(\ref{eq:dJarl}), one can immediately observe that the first and second lines on the right-hand side of Eq.~(\ref{eq:dJarl}) become
\begin{eqnarray}
{\rm Im} \left[\dot{V}^{}_{e1} V^{}_{\mu 2} V^*_{e2} V^*_{\mu 1}\right] + {\rm Im} \left[V^{}_{e1} V^{}_{\mu 2} V^*_{e2} \dot{V}^*_{\mu 1}\right] &=& \widetilde{\cal J} \left[+|V^{}_{e2}|^2 \widetilde{\Delta}^{-1}_{12} - \left(|V^{}_{e3}|^2 - |V^{}_{e1}|^2\right)\widetilde{\Delta}^{-1}_{31}\right]\; , \nonumber \\
{\rm Im} \left[V^{}_{e1} \dot{V}^{}_{\mu 2} V^*_{e2} V^*_{\mu 1}\right] + {\rm Im} \left[V^{}_{e1} V^{}_{\mu 2} \dot{V}^*_{e2} V^*_{\mu 1}\right] &=& \widetilde{\cal J} \left[ - |V^{}_{e1}|^2 \widetilde{\Delta}^{-1}_{12} - \left(|V^{}_{e2}|^2 - |V^{}_{e3}|^2\right)\widetilde{\Delta}^{-1}_{23}\right]\; , ~~~~~
\end{eqnarray}
leading to the following simple result
\begin{eqnarray}\label{eq:jarlskog}
\frac{{\rm d}}{{\rm d}a} \widetilde{\cal J} = - \widetilde{\cal J}
\left[\left(|V^{}_{e1}|^2 - |V^{}_{e2}|^2\right) \widetilde{\Delta}^{-1}_{12} + \left(|V^{}_{e2}|^2 - |V^{}_{e3}|^2\right) \widetilde{\Delta}^{-1}_{23} + \left(|V^{}_{e3}|^2 - |V^{}_{e1}|^2\right) \widetilde{\Delta}^{-1}_{31} \right] \; .
\end{eqnarray}
Combining Eq.~(\ref{eq:dDeltaij}) and Eq.~(\ref{eq:jarlskog}), one arrives at
\begin{eqnarray}
\frac{{\rm d}}{{\rm d}a} \ln \left[\widetilde{\cal J} \widetilde{\Delta}^{}_{12} \widetilde{\Delta}^{}_{23} \widetilde{\Delta}^{}_{31}\right] = 0 \; ,
\end{eqnarray}
implying the well-known Naumov relation~\cite{Naumov:1991ju}. The corresponding identity $\widetilde{\cal J}\widetilde{\Delta}^{}_{12} \widetilde{\Delta}^{}_{23} \widetilde{\Delta}^{}_{31} = {\cal J} \Delta^{}_{12} \Delta^{}_{23} \Delta^{}_{31}$ has previously been derived in the literature by implementing the commutators of effective lepton mass matrices~\cite{Harrison:1999df, Xing:2000gg}. In addition to the Naumov relation, it is easy to verify that $\displaystyle\sum_i \widetilde{m}^2_i V^*_{\alpha i} V^{}_{\beta i} = \sum_i m^2_i U^*_{\alpha i} U^{}_{\beta i}$ holds for arbitrary $\alpha$ and $\beta$ except for $\alpha = \beta = e$.

\item For completeness, we explicitly write down the RGEs of $|V^{}_{\mu i}|^2$, which can also be expressed in terms of $|V^{}_{\alpha i}|^2$ and $\widetilde{\Delta}^{}_{ij}$. Based on Eq.~(\ref{eq:Vai2}) and the results from Ref.~\cite{Chiu:2017ckv}, one can find
\begin{eqnarray}\label{eq:Vmi2}
\frac{\rm d}{{\rm d}a} |V^{}_{\mu 1}|^2 &=& |V^{}_{\mu 1}|^2 \left[\frac{|V^{}_{e2}|^2}{\widetilde{\Delta}^{}_{12}} - \frac{|V^{}_{e3}|^2}{\widetilde{\Delta}^{}_{31}}\right] + |V^{}_{e 1}|^2 \left[\frac{|V^{}_{\mu 2}|^2}{\widetilde{\Delta}^{}_{12}} - \frac{|V^{}_{\mu 3}|^2}{\widetilde{\Delta}^{}_{31}}\right] - \left[\frac{|V^{}_{\tau 3}|^2}{\widetilde{\Delta}^{}_{12}} - \frac{|V^{}_{\tau 2}|^2}{\widetilde{\Delta}^{}_{31}}\right]\; , \nonumber \\
\frac{\rm d}{{\rm d}a} |V^{}_{\mu 2}|^2 &=& |V^{}_{\mu 2}|^2 \left[\frac{|V^{}_{e3}|^2}{\widetilde{\Delta}^{}_{23}} - \frac{|V^{}_{e1}|^2}{\widetilde{\Delta}^{}_{12}}\right] + |V^{}_{e 2}|^2 \left[\frac{|V^{}_{\mu 3}|^2}{\widetilde{\Delta}^{}_{23}} - \frac{|V^{}_{\mu 1}|^2}{\widetilde{\Delta}^{}_{12}}\right] - \left[\frac{|V^{}_{\tau 1}|^2}{\widetilde{\Delta}^{}_{23}} - \frac{|V^{}_{\tau 3}|^2}{\widetilde{\Delta}^{}_{12}}\right] \; , \nonumber \\
\frac{\rm d}{{\rm d}a} |V^{}_{\mu 3}|^2 &=& |V^{}_{\mu 3}|^2 \left[\frac{|V^{}_{e1}|^2}{\widetilde{\Delta}^{}_{31}} - \frac{|V^{}_{e2}|^2}{\widetilde{\Delta}^{}_{23}}\right] + |V^{}_{e 3}|^2 \left[\frac{|V^{}_{\mu 1}|^2}{\widetilde{\Delta}^{}_{31}} - \frac{|V^{}_{\mu 2}|^2}{\widetilde{\Delta}^{}_{23}}\right] - \left[\frac{|V^{}_{\tau 2}|^2}{\widetilde{\Delta}^{}_{31}} - \frac{|V^{}_{\tau 1}|^2}{\widetilde{\Delta}^{}_{23}}\right] \; .
\end{eqnarray}
The RGEs of $|V^{}_{\tau i}|^2$ can be obtained from Eq.~(\ref{eq:Vmi2}) by simply exchanging $|V^{}_{\mu i}|^2$ with $|V^{}_{\tau i}|^2$ for $i = 1, 2, 3$. It is now evident that the evolution of $|V^{}_{\mu i}|^2$ (or $|V^{}_{\tau i}|^2$) is governed not only by $|V^{}_{\mu i}|^2$ (or $|V^{}_{\tau i}|^2$) and $\widetilde{\Delta}^{}_{ij}$, but also by $|V^{}_{\tau i}|^2$ (or $|V^{}_{\mu i}|^2$) and $|V^{}_{e i}|^2$. Comparing between Eq.~(\ref{eq:Vei2}) and Eq.~(\ref{eq:Vmi2}), one can easily notice the special role played by the electron flavor in studying the matter effects on the neutrino flavor mixing parameters.
\end{itemize}
\begin{figure}[!t]
\centering
\includegraphics[width=\textwidth]{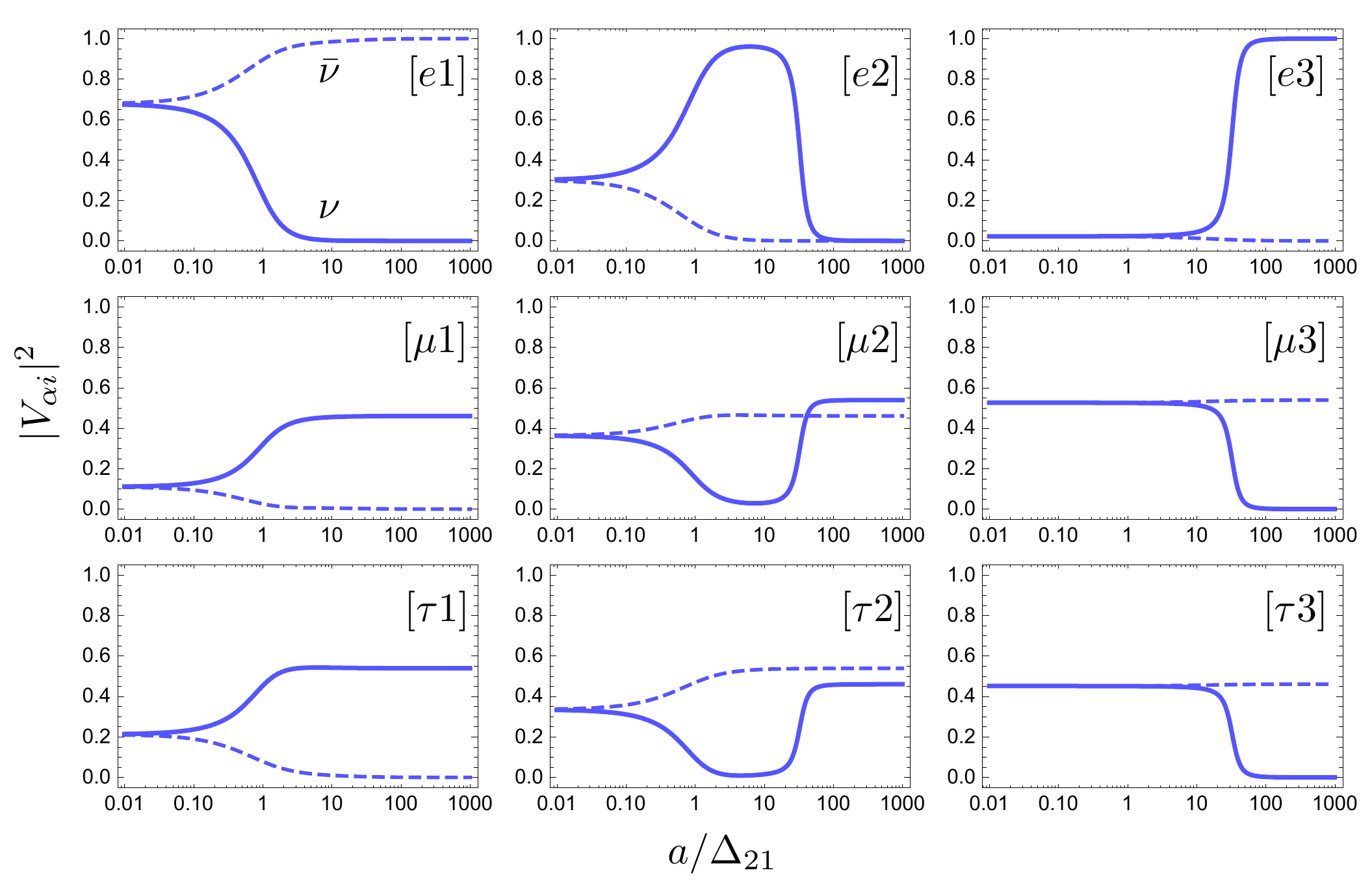}
\vspace{-1.0cm}
\caption{\label{fig:mixing_NO} The evolution of the effective mixing matrix elements in matter $|V^{}_{\alpha i}|^2$ (for $\alpha = e, \mu, \tau$ and $i = 1, 2, 3$) with respect to the parameter $a/\Delta^{}_{21}$, where the best-fit values of neutrino mixing parameters from Ref.~\cite{Esteban:2016qun} in the NO case are input and the blue solid (dashed) curves correspond to the results of neutrino (antineutrino) oscillations. }
\end{figure}
\begin{figure}[!t]
\centering
\includegraphics[width=\textwidth]{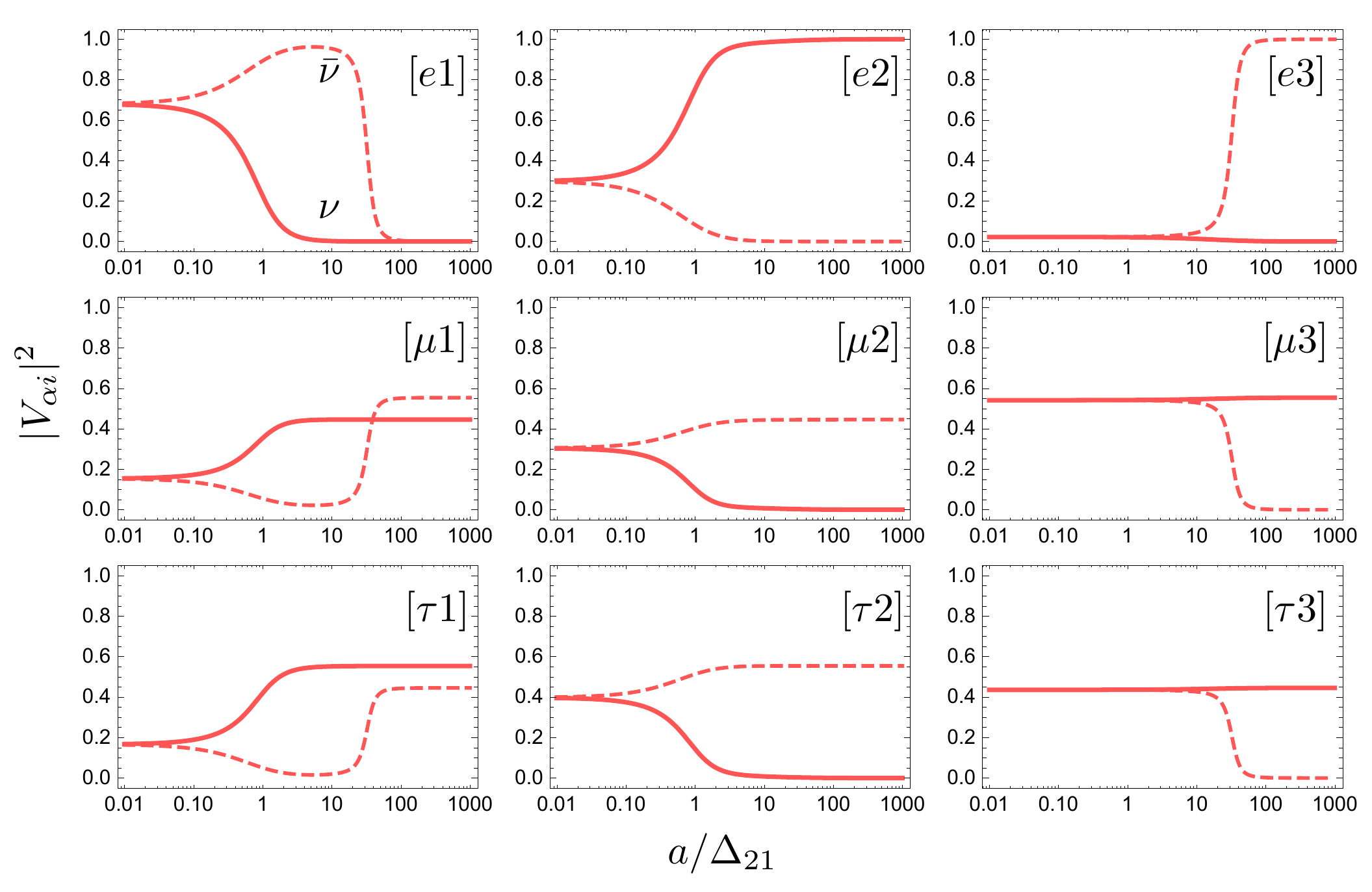}
\vspace{-1.0cm}
\caption{\label{fig:mixing_IO}  The evolution of the effective mixing matrix elements in matter $|V^{}_{\alpha i}|^2$ (for $\alpha = e, \mu, \tau$ and $i = 1, 2, 3$) with respect to the matter term $a/\Delta^{}_{21}$, where the best-fit values of neutrino mixing parameters from Ref.~\cite{Esteban:2016qun} in the IO case are input and the red solid (dashed) curves correspond to the results of neutrino (antineutrino) oscillations. }
\end{figure}

The central results for the RGEs of the leptonic flavor mixing matrix $V$ and neutrino masses $\widetilde{m}^{}_i$ in matter are given in Eqs.~(\ref{eq:diag}) and (\ref{eq:Vai2}). For illustration, we show the evolution of $|V^{}_{\alpha i}|^2$ against the dimensionless parameter $a/\Delta^{}_{21}$ in Figs. \ref{fig:mixing_NO} and \ref{fig:mixing_IO}, where the best-fit values of all the neutrino mixing parameters from Ref.~\cite{Esteban:2016qun} have been used in our numerical calculations. Throughout this paper, the blue solid (dashed) curves are referred to the results for neutrino (antineutrino) oscillations in the NO case, whereas the red solid (dashed) curves to those for neutrino (antineutrino) oscillations in the IO case. The main features of the evolution of $|V^{}_{\alpha i}|^2$ can be understood by using the RGEs in Eqs.~(\ref{eq:Vei2}) and (\ref{eq:dDeltaij}) together with the general properties of matter effects themselves:
\begin{enumerate}
\item First of all, it should be stressed that the evolution of $|V^{}_{\mu i}|^2$ is qualitatively identical to that of $|V^{}_{\tau i}|^2$ for $i = 1, 2, 3$, comparing the plots in the second row and those in the third row of Fig.~1. This behavior can be well understood by noticing that the muon and tau flavors are indistinguishable, since muon and tau neutrinos (antineutrinos) experience only the universal neutral-current interactions in ordinary matter. In addition, the initial values of $|V^{}_{\mu i}|^2$ and $|V^{}_{\tau i}|^2$ at $a = 0$, namely, the mixing matrix elements in vacuum, approximately respect the $\mu$-$\tau$ symmetry $|U^{}_{\mu i}|^2 = |U^{}_{\tau i}|^2$ for $i = 1, 2, 3$. The slight breaking of this symmetry will be responsible for the quantitative difference between the evolution of $|V^{}_{\mu i}|^2$ and that of $|V^{}_{\tau i}|^2$. This conclusion is also applicable to antineutrinos. For this reason, we shall only concentrate on the electron and muon flavors.

\item As the matrix elements have to fulfill the unitarity condition $|V^{}_{e i}|^2 + |V^{}_{\mu i}|^2 + |V^{}_{\tau i}|^2 = 1$ for $i = 1, 2, 3$, it is then necessary to consider only $|V^{}_{e i}|^2$ in the first row of Fig.~1. First, the evolution of $|V^{}_{e 1}|^2$ is governed by the first equation in Eq.~(\ref{eq:Vei2}). At the beginning, we have $\widetilde{\Delta}^{}_{12} = - \Delta^{}_{21} < 0$ and $\widetilde{\Delta}^{}_{31} = \Delta^{}_{31} \gg \Delta^{}_{21}$, so the derivative of $|V^{}_{e1}|^2$ is approximately given by $-2 |U^{}_{e1}|^2 \Delta^{-1}_{21} < 0$, indicating that $|V^{}_{e1}|^2$ decreases with the increasing $a$. Similarly, one can observe from the second equation in Eq.~(\ref{eq:Vei2}) that $|V^{}_{e2}|^2$ is increasing. On the other hand, at the early stage, the evolution of $|V^{}_{e3}|^2$ is highly suppressed by both $|V^{}_{e3}|^2 = |U^{}_{e3}|^2$ itself and the large neutrino mass-squared difference $\Delta^{}_{31} \approx \Delta^{}_{32}$, as one can see from the third equation of Eq.~(\ref{eq:Vei2}). Then, the resonance corresponding to $\Delta^{}_{21}$ is reached around $a/\Delta^{}_{21} = 1$, where $|V^{}_{e1}|^2 = |V^{}_{e2}|^2$ is satisfied and the changing rates of $|V^{}_{e1}|^2$ and $|V^{}_{e2}|^2$ maximize. Looking at again the RGE of $|V^{}_{e2}|^2$, we find that as $|V^{}_{e1}|^2$ decreases and $\widetilde{\Delta}^{}_{21}$ increases, the right-hand side of the second equation of Eq.~(\ref{eq:Vei2}) first approaches zero and then changes its sign. This means that $|V^{}_{e2}|^2$ reaches its maximum and decreases to zero afterwards. The decreasing rate is maximal around the MSW resonance corresponding to $\Delta^{}_{31}$, namely, $a/\Delta^{}_{21} \approx \Delta^{}_{31}/\Delta^{}_{21} \approx 30$. Finally, since both $|V^{}_{e1}|^2$ and $|V^{}_{e2}|^2$ become vanishing for an extremely large $a$, we have $|V^{}_{e3}|^2$ close to one due to the unitarity condition $|V^{}_{e1}|^2 + |V^{}_{e2}|^2 + |V^{}_{e3}|^2 = 1$.

\item Now we consider the results for antineutrinos in the NO case, as represented by the dashed curves in Fig.~1. It is worth emphasizing that the replacements $U \to U^*$ and $a \to -a$ have been made and thus the matter term $a$ itself keeps positive for both neutrinos and antineutrinos. As a consequence, the right-hand sides of the RGEs in Eqs.~(\ref{eq:Vei2}) and (\ref{eq:dDeltaij}) should be multiplied by a negative sign when applied to antineutrinos. Furthermore, in the NO case, there are no MSW resonances for antineutrinos, so the evolution of $|V^{}_{e1}|^2$ and $|V^{}_{e2}|^2$ seems to be milder and in the opposite directions, compared with the results for neutrinos. In particular, $|V^{}_{e3}|^2$ is monotonically decreasing from the initial value to zero in the end.
\end{enumerate}
The numerical results for neutrinos and antineutrinos in the IO case have been given in Fig.~2. One can analyze the evolution of $|V^{}_{ei}|^2$ in a very similar way to the NO case. The difference between these two cases is the location of the MSW resonances. As is well known, the $\Delta^{}_{21}$-driven resonance remains for neutrinos in the IO case, while the $\Delta^{}_{31}$-driven resonance is absent. But the opposite is true for antineutrinos. Bearing these general features in mind, one can easily understand the behaviors of $|V^{}_{\alpha i}|^2$ evolving with an increasing $a/\Delta^{}_{21}$.

\begin{figure}[!t]
\centering
\includegraphics[width=0.8\textwidth]{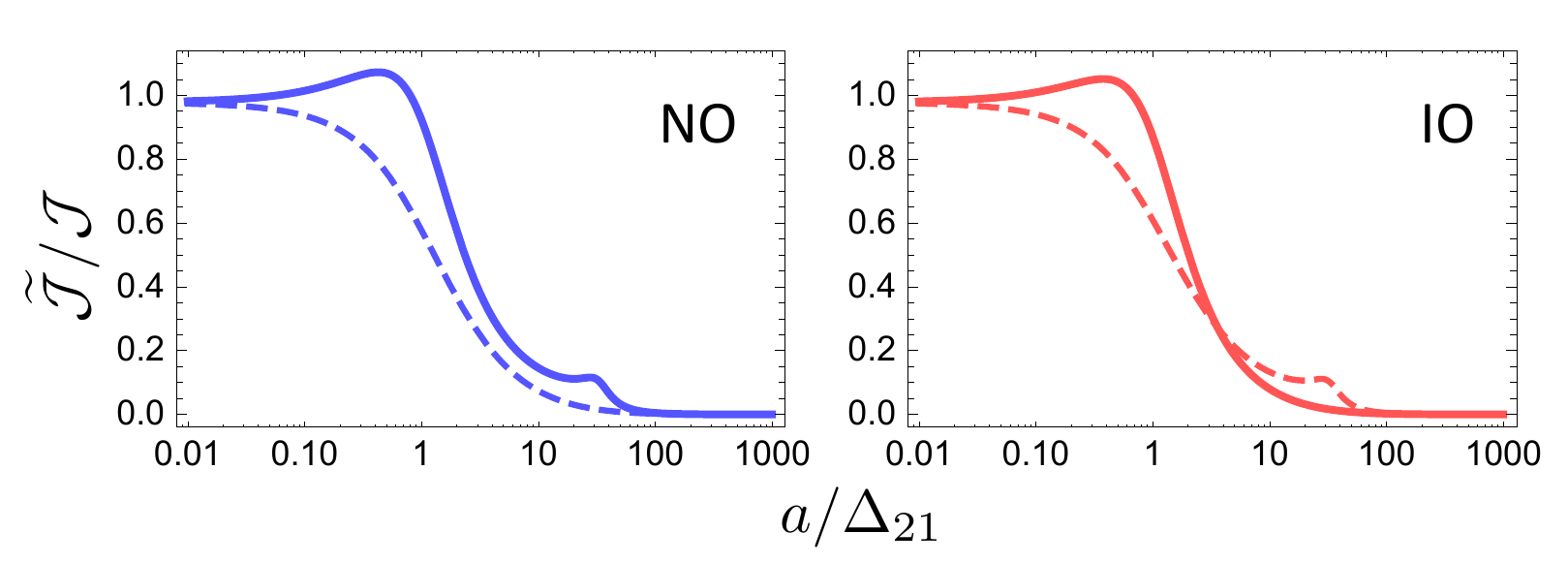}
\vspace{-0.3cm}
\caption{\label{fig:Jarlskog}  The evolution of the Jarlskog invariant $\widetilde{\mathcal{J}}$, normalized by its vacuum value $\mathcal{J}$,  with respect to $a/\Delta^{}_{21}$. The numerical results in the NO and IO cases are given in the left and right panel, respectively, where the same input values as before are adopted. }
\end{figure}

The running behavior of the Jarlskog invariant $\widetilde{\cal J}$, normalized by its vacuum value ${\cal J}$, is given in Fig.~\ref{fig:Jarlskog} in both NO and IO cases. For neutrinos, as we have mentioned, both $\Delta^{}_{21}$- and $\Delta^{}_{31}$-driven resonances take place in the NO case, corresponding to two local maxima of $\widetilde{\cal J}/{\cal J}$. The existence of these two maxima can be partly understood by examining the right-hand side of Eq.~(\ref{eq:jarlskog}). In the early stage of evolution, e.g., $a \lesssim \Delta^{}_{21} \ll \Delta^{}_{31}$, one can safely assume $\widetilde{\Delta}^{}_{32} = \widetilde{\Delta}^{}_{31} \gg \widetilde{\Delta}^{}_{21}$ and thus ignore the last two terms. This leads to ${\rm d}(\ln \widetilde{\cal J})/{\rm d}a \approx (|V^{}_{e1}|^2 - |V^{}_{e2}|^2)/\widetilde{\Delta}^{}_{21}$, which is vanishingly small at the resonance around $a/\Delta^{}_{21} \approx 1$ and $|V^{}_{e1}|^2 \approx |V^{}_{e2}|^2$. On the other hand, when $a/\Delta^{}_{21} \sim 10$, one can read from the first row of Fig.~1 that $|V^{}_{e1}|^2 \approx 0$ and $|V^{}_{e2}|^2 \approx 1 - |V^{}_{e3}|^2$. The term $(|V^{}_{e 2}|^2 - |V^{}_{e3}|^2)/\widetilde{\Delta}^{}_{32}$ becomes dominant at the late stage as $\widetilde{\Delta}^{}_{32} \to \Delta^{}_{21}$ and $\widetilde{\Delta}^{}_{21} \approx \widetilde{\Delta}^{}_{31}$, so the second maximum of $\widetilde{\cal J}$ is obtained at the $\Delta^{}_{31}$-driven resonance with $|V^{}_{e2}|^2 \approx |V^{}_{e3}|^2$. However, the explicit expressions of $|V^{}_{ei}|^2$ and $\widetilde{\Delta}^{}_{ij}$ are needed to figure out the exact values of $a$ for the local maxima. See, e.g., Ref.~\cite{Xing:2016ymg}, for the discussions about the first local maximum. The numerical results in the IO case or those for antineutrinos in both cases can be understood by studying the appearance of the MSW resonances.

\section{Mixing Angles and CP-violating Phase}

Although the RGEs for the mixing matrix elements $V^{}_{\alpha i}$ are sufficient to explore their dependence on the matter term $a$, it will be instructive to derive the RGEs for the effective mixing angles $\{\widetilde{\theta}^{}_{12}, \widetilde{\theta}^{}_{13}, \widetilde{\theta}^{}_{23}\}$ and the CP-violating phase $\widetilde{\delta}$ in the standard parametrization~\cite{Patrignani:2016xqp}. The motivation for such an investigation is two-fold. First of all, the RGEs of neutrino mixing parameters due to radiative corrections have been extensively studied~\cite{Ohlsson:2013xva}. A detailed comparison between the RGEs arising from quantum corrections and those from matter effects in neutrino oscillations will be very helpful. Second, as neutrino oscillation behaviors are usually understood in terms of neutrino mixing parameters and neutrino mass-squared differences, the impact of matter effects on neutrino oscillations can be conveniently represented by the effective flavor mixing angles and CP-violating phase in matter.

The effective mixing matrix $V$, which is a $3\times 3$ unitary matrix, can in general be parametrized in terms of three mixing angles and six phases, namely, $V = Q\cdot U^\prime \cdot P$ with $Q \equiv {\rm diag}\{e^{{\rm i}\varphi^{}_1}, e^{{\rm i}\varphi^{}_2}, 1\}$ and $P \equiv {\rm diag}\{e^{{\rm i}\phi^{}_1}, e^{{\rm i}\phi^{}_2}, e^{{\rm i}\phi^{}_3}\}$. The unitary matrix $U^\prime$ takes on the same form as in Eq.~(\ref{eq:PMNS}) with the mixing angles and CP-violating phase replaced by $\{\widetilde{\theta}^{}_{12}, \widetilde{\theta}^{}_{13}, \widetilde{\theta}^{}_{23}\}$ and $\widetilde{\delta}$. As $V D V^\dagger = Q\cdot U^\prime D U^{\prime \dagger} \cdot Q^\dagger$, it is obvious that the diagonal phase matrix $P$ disappears from Eq.~(\ref{eq:RGE}). Therefore, we can just ignore $P$, but have to retain $Q$, in which two unphysical phases $\varphi^{}_1$ and $\varphi^{}_2$ are involved. Taking $V = Q U^\prime$ and noticing $V^\dagger \dot{V} = U^{\prime \dagger} \left( Q^\dagger \dot{Q} \right) U^\prime + U^{\prime \dagger} \dot{U}^\prime$, we arrive at
\begin{eqnarray}\label{eq:rgepara}
\sum_\alpha U^{\prime *}_{\alpha i} \dot{U}^\prime_{\alpha j} + {\rm i} \left[\dot{\varphi}^{}_1 U^{\prime *}_{ei} U^{\prime}_{ej} + \dot{\varphi}^{}_2 U^{\prime *}_{\mu i} U^{\prime}_{\mu j} \right] = U^{\prime *}_{e i} U^{\prime}_{ej} \widetilde{\Delta}^{-1}_{ji} \; ,
\end{eqnarray}
for $ij = 12, 13, 23$. The diagonal elements give rise to $\dot{\widetilde{m}^2_i} = |U^\prime_{ei}|^2$ as before. It is worthwhile to stress that Eq.~(\ref{eq:rgepara}) resembles the salient features of the ordinary RGEs for quantum corrections to the lepton flavor mixing parameters in the case of massive Dirac neutrinos, particularly in the limit of so-called tau-lepton dominance~\cite{Xing:2005fw}. In comparison with the tau-lepton dominance due to $y^2_e \ll y^2_\mu \ll y^2_\tau$, where $y^{}_\alpha$ (for $\alpha = e, \mu, \tau$) stand for the charged-lepton Yukawa couplings, the case of matter effects under consideration corresponds to the electron dominance, since the coherent forward scattering of neutrinos in the normal matter singles out the electron flavor. As a consequence, the standard parametrization and the original Kobayashi-Maskawa parametrization~\cite{Kobayashi:1973fv, Zhou:2011xm} with the simplest matrix elements in the first row will be most convenient for us to derive the RGEs of relevant flavor mixing parameters.

Adopting the standard parametrization of $U^\prime$, we get the equation array for the derivatives of the flavor mixing parameters $\{\dot{\widetilde{\theta}}^{}_{12}, \dot{\widetilde{\theta}}^{}_{13}, \dot{\widetilde{\theta}}^{}_{23}, \dot{\widetilde{\delta}}, \dot{\varphi}^{}_1, \dot{\varphi}^{}_2\}$ from both imaginary and real parts of Eq.~(\ref{eq:rgepara}) for $ij = 12, 13, 23$. After a lengthy but straightforward calculation, we find the RGEs for four physical mixing parameters
\begin{eqnarray}\label{eq:anglephase}
\dot{\widetilde{\theta}}^{}_{12} &=& \frac{1}{2} \sin 2\widetilde{\theta}^{}_{12} \left(\cos^2 \widetilde{\theta}^{}_{13} \widetilde{\Delta}^{-1}_{21} - \sin^2 \widetilde{\theta}^{}_{13} \widetilde{\Delta}^{}_{21} \widetilde{\Delta}^{-1}_{31} \widetilde{\Delta}^{-1}_{32} \right) \; , \nonumber \\
\dot{\widetilde{\theta}}^{}_{13} &=& \frac{1}{2} \sin 2\widetilde{\theta}^{}_{13} \left(\cos^2 \widetilde{\theta}^{}_{12} \widetilde{\Delta}^{-1}_{31} + \sin^2 \widetilde{\theta}^{}_{12} \widetilde{\Delta}^{-1}_{32} \right) \; , \nonumber \\
\dot{\widetilde{\theta}}^{}_{23} &=& \frac{1}{2} \sin 2\widetilde{\theta}^{}_{12} \sin \widetilde{\theta}^{}_{13} \cos \widetilde{\delta} \widetilde{\Delta}^{}_{21} \widetilde{\Delta}^{-1}_{31} \widetilde{\Delta}^{-1}_{32} \; , \nonumber \\
\dot{\widetilde{\delta}} &=& - \sin 2\widetilde{\theta}^{}_{12} \sin \widetilde{\theta}^{}_{13} \sin \widetilde{\delta} \cot 2\widetilde{\theta}^{}_{23} \widetilde{\Delta}^{}_{21} \widetilde{\Delta}^{-1}_{31} \widetilde{\Delta}^{-1}_{32} \; ;
\end{eqnarray}
and those for two unphysical phases
\begin{eqnarray}\label{eq:unphases}
\dot{\varphi}^{}_1 &=& - \frac{1}{2} \sin 2\widetilde{\theta}^{}_{12} \sin \widetilde{\theta}^{}_{13} \sin \widetilde{\delta} \tan \widetilde{\theta}^{}_{23} \widetilde{\Delta}^{}_{21} \widetilde{\Delta}^{-1}_{31} \widetilde{\Delta}^{-1}_{32} \; , \nonumber \\
\dot{\varphi}^{}_2 &=& - \sin 2\widetilde{\theta}^{}_{12} \sin \widetilde{\theta}^{}_{13} \sin \widetilde{\delta} \csc 2\widetilde{\theta}^{}_{23} \widetilde{\Delta}^{}_{21} \widetilde{\Delta}^{-1}_{31} \widetilde{\Delta}^{-1}_{32} \; .
\end{eqnarray}
Using the last two equations in Eq.~(\ref{eq:anglephase}), one can easily verify
\begin{eqnarray}
\frac{{\rm d}}{{\rm d}a} \left(\sin 2\widetilde{\theta}^{}_{23} \sin \widetilde{\delta}\right) = \left(2\cos 2\widetilde{\theta}^{}_{23} \sin \widetilde{\delta}\right) \dot{\widetilde{\theta}}^{}_{23} + \left(\sin 2\widetilde{\theta}^{}_{23} \cos\widetilde{\delta} \right) \dot{\widetilde{\delta}} = 0 \; ,
\end{eqnarray}
which is just the Toshev relation $\sin 2\widetilde{\theta}^{}_{23} \sin \widetilde{\delta} = \sin 2\theta^{}_{23} \sin \delta$ in the standard parametrization. Some comments in the RGEs in Eqs.~(\ref{eq:anglephase}) and (\ref{eq:unphases}) are in order.
\begin{figure}[!t]
\centering
\includegraphics[width=0.8\textwidth]{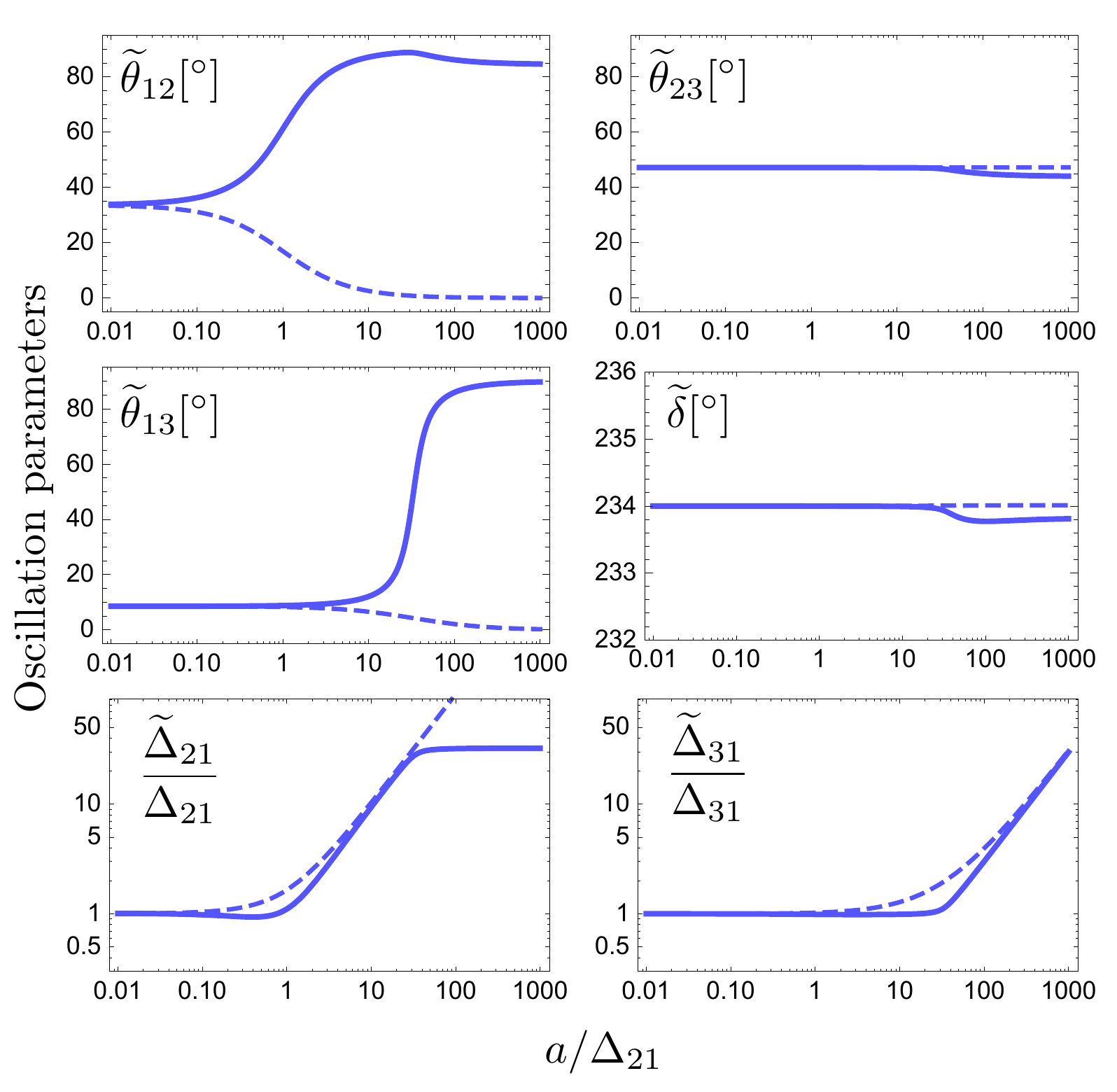}
\vspace{-0.4cm}
\caption{\label{fig:osci_para_NO} The evolution of the effective flavor mixing parameters in matter, namely, three mixing angles $\{\widetilde{\theta}^{}_{12}, \widetilde{\theta}^{}_{13}, \widetilde{\theta}^{}_{23}\}$, one CP-violating phase $\widetilde{\delta}$ and two mass-squared differences $\{\widetilde{\Delta}_{21},\widetilde{\Delta}_{31}\}$, with respect to $a/\Delta^{}_{21}$ in the NO case. The same convention and input values as in Fig.~\ref{fig:mixing_NO} are taken. }
\end{figure}
\begin{itemize}
\item If there exists a $\mu$-$\tau$ symmetry in the lepton flavor mixing matrix $U$ in vacuum, namely, $|U^{}_{\mu i}|^2 = |U^{}_{\tau i}|^2$ for $i = 1, 2, 3$, the mixing parameters should satisfy $\theta^{}_{23} = \pi/4$ and $\delta = \pm \pi/2$. As has been proved in Ref.~\cite{Xing:2015fdg}, the matter effects preserve the $\mu$-$\tau$ symmetry $|V^{}_{\mu i}|^2 = |V^{}_{\tau i}|^2$ (for $i = 1, 2, 3$), i.e., $\widetilde{\theta}^{}_{23} = \pi/4$ and $\widetilde{\delta} = \pm \pi/2$. This conclusion can be understood via the RGEs of $\widetilde{\theta}^{}_{23}$ and $\widetilde{\delta}$ in Eq.~(\ref{eq:anglephase}). For instance, the initial conditions $\widetilde{\theta}^{}_{23}|^{}_{a = 0} = \theta^{}_{23} = \pi/4$ and $\widetilde{\delta}|^{}_{a = 0} = \delta = \pm \pi/2$ guarantee that the beta functions of $\dot{\widetilde{\theta}}^{}_{23} \propto \cos \widetilde{\delta}$ and $\dot{\widetilde{\delta}} \propto \cos 2\widetilde{\theta}^{}_{23}$ are vanishing, indicating that the $\mu$-$\tau$ symmetry with $\widetilde{\theta}^{}_{23} = \pi/4$ and $\widetilde{\delta} = \pm \pi/2$ is fully stable against matter effects.

\item Furthermore, let us look for possible fixed points of other flavor mixing parameters in their running with $a$. First, starting from the mixing angles and CP-violating phase in vacuum, we can see that $\dot{\widetilde{\theta}}^{}_{12} \approx \sin 2\widetilde{\theta}^{}_{12} \cos^2 \widetilde{\theta}^{}_{13} \widetilde{\Delta}^{-1}_{21}/2$ is positive, where the other term $\tan^2 \widetilde{\theta}^{}_{13} \widetilde{\Delta}^2_{21}/(\widetilde{\Delta}^{}_{31} \widetilde{\Delta}^{}_{32}) \ll 1$ at $a = 0$ has been neglected. Therefore, $\widetilde{\theta}^{}_{12}$ increases as the matter density or neutrino energy becomes larger. Second, if $\theta^{}_{13} = 0$ is assumed, then one can observe $\dot{\widetilde{\theta}}^{}_{13} = 0$ and $\dot{\widetilde{\theta}}^{}_{23} = 0$. In this case, only the mixing angle $\widetilde{\theta}^{}_{12}$ will be affected by matter effects, and the CP-violating phase $\widetilde{\delta}$ is not well-defined and thus irrelevant. Third, we assume $\delta = 0$ or $\pi$ in vacuum, corresponding to the case of CP conservation, and then obtain $\dot{\widetilde{\delta}} \propto \sin \widetilde{\delta} = 0$, implying that the CP-violating phase $\widetilde{\delta}$ is fixed. However, the CP asymmetries between neutrino and antineutrino oscillation probabilities could exisit since the mass-squared differences for neutrinos and antineutrinos will be different due to the opposite signs in front of $a$. This is just the fake CP violation induced by matter effects.
\end{itemize}
\begin{figure}[!t]
\centering
\includegraphics[width=0.8\textwidth]{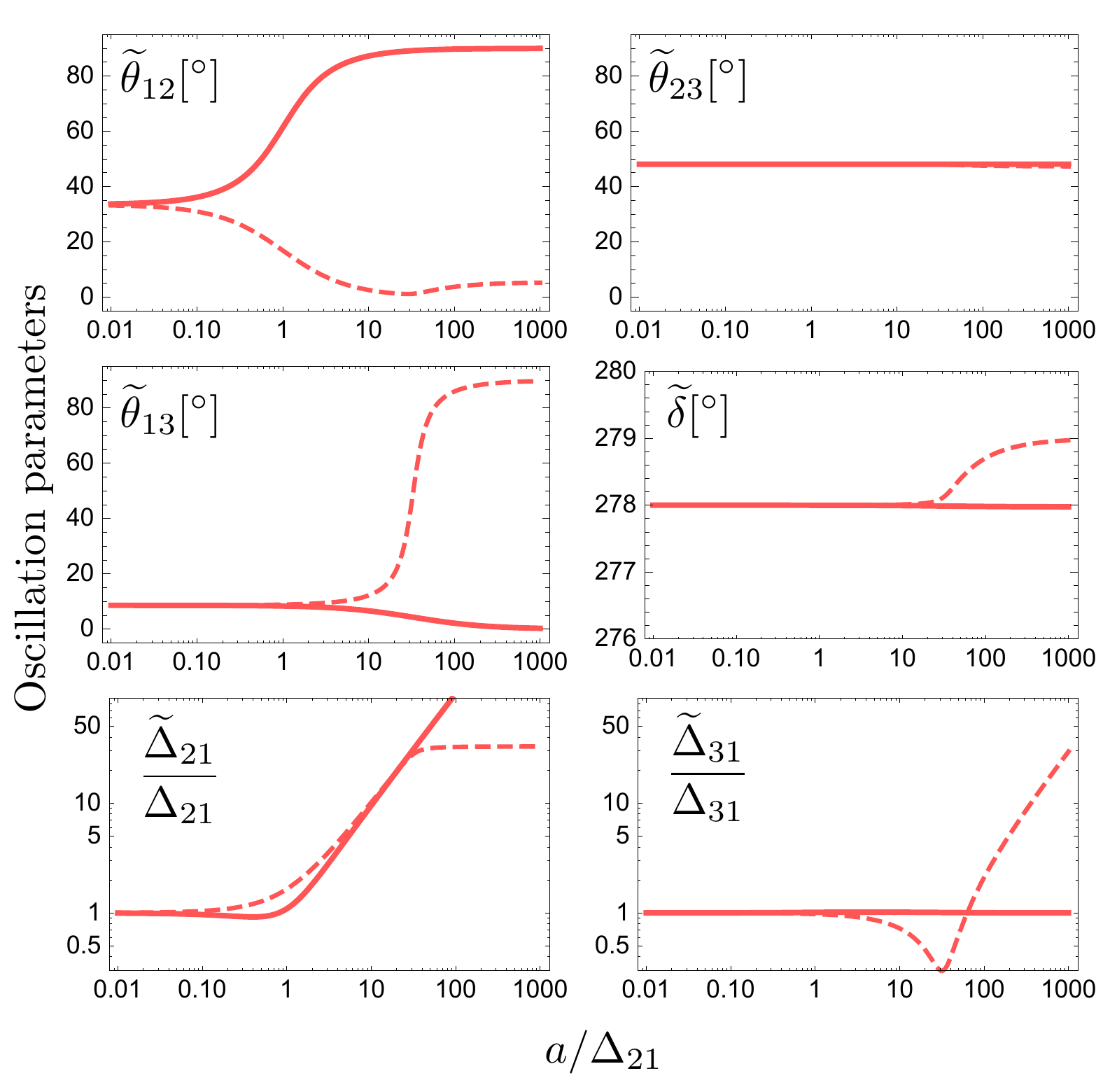}
\vspace{-0.4cm}
\caption{\label{fig:osci_para_IO} The evolution of the effective flavor mixing parameters in matter, namely, three mixing angles $\{\widetilde{\theta}^{}_{12}, \widetilde{\theta}^{}_{13}, \widetilde{\theta}^{}_{23}\}$, one CP-violating phase $\widetilde{\delta}$ and two mass-squared differences $\{\widetilde{\Delta}_{21},\widetilde{\Delta}_{31}\}$, with respect to $a/\Delta^{}_{21}$ in the IO case. The same convention and input values as in Fig.~\ref{fig:mixing_IO} are taken. }
\end{figure}

We can numerically solve the RGEs of the mixing parameters $\{\widetilde{\theta}^{}_{12}, \widetilde{\theta}^{}_{13}, \widetilde{\theta}^{}_{23}, \widetilde{\delta}\}$ and the neutrino mass-squared differences $\{\widetilde{\Delta}^{}_{21}, \widetilde{\Delta}^{}_{31}\}$. However, as the evolution of $|V^{}_{\alpha i}|^2$ (for $\alpha = e, \mu, \tau$ and $i = 1, 2, 3$) have been obtained, we extract the results of mixing parameters from the calculations of $|V^{}_{\alpha i}|^2$ for Figs.~1 and 2 and summarize them in Figs.~\ref{fig:osci_para_NO} and \ref{fig:osci_para_IO}, where the input values are the same as before. The running behaviors of $\widetilde{\theta}^{}_{12}$ and $\widetilde{\theta}^{}_{13}$ are directly extracted from those of $|V^{}_{e2}|^2 = \cos^2 \widetilde{\theta}^{}_{13} \cos^2 \widetilde{\theta}^{}_{12}$ and $|V^{}_{e3}|^2 = \sin^2 \widetilde{\theta}^{}_{13}$ in the standard parametrization. For instance, we have $\tan^2 \widetilde{\theta}^{}_{12} = |V^{}_{e2}|^2/|V^{}_{e1}|^2$. Note that $\widetilde{\theta}^{}_{12} \to 0$ or $90^\circ$ after crossing the first MSW resonance, while $\widetilde{\theta}^{}_{13} \to 0$ or $90^\circ$ after crossing the second resonance no matter how small the value of $\theta^{}_{13}$ in vacuum is. As for $\widetilde{\theta}^{}_{23}$ and $\widetilde{\delta}$, since current neutrino oscillation data prefer nearly-maximal mixing angle and CP-violating phase, the matter effects have very little influence on their values in matter, which is well consistent with the Toshev relation.

As shown in Figs.~4 and 5, all the effective mixing angles become constant in the limit $a/\Delta^{}_{21} \to +\infty$. In other words, the infinity serves as a special fixed point of the RGEs of mixing angles. In both NO and IO cases, one can observe that $\widetilde{\theta}^{}_{13}$ always approaches either $0$ or $90^\circ$ for both neutrinos and antineutrinos in this limit. These values are asymptotically stable because $\dot{\widetilde{\theta}}^{}_{13}$ vanishes at either $\widetilde{\theta}^{}_{13} = 0$ or $90^\circ$. However, the
limits of $\widetilde{\theta}^{}_{12}$ and $\widetilde{\theta}^{}_{23}$ depend upon the asymptotic value of $\widetilde{\theta}^{}_{13}$. Following the perturbation calculations in Refs.~\cite{Cervera:2000kp, Freund:2001pn} in the limit of $\Delta^{}_{21} \ll \Delta^{}_{31} \ll a$, we obtain
\begin{eqnarray}
\cot\widetilde{\theta}_{12} \to \frac{\Delta^{}_{21}}{\Delta^{}_{31}}
\cdot\frac{c^{}_{12} s^{}_{12} }{c_{13}^2 s^{}_{13}} \; , \quad
\tan\widetilde{\theta}_{23} \to \left| \tan\theta_{23} + e^{{\rm i}\delta}
\frac{\Delta_{21}}{\Delta_{31}} \cdot\frac{c_{12} s_{12} }{c_{23}^2 s_{13}}
\right| \; ,
\end{eqnarray}
for neutrinos in the NO case. For antineutrinos in the IO case, the results can be obtained by replacing $\cot\widetilde{\theta}_{12}$ by
$\tan\widetilde{\theta}_{12}$ but keeping $\tan\widetilde{\theta}_{23}$ unchanged. In the NO case for neutrinos, with the best-fit values of the mixing angles and the CP-violating phase in vacuum, one can figure out the asymptotic values $\widetilde{\theta}^{}_{12} \approx 84.6^\circ$ and $\widetilde{\theta}^{}_{23} \approx 44.3^\circ$ in the limit $a/\Delta^{}_{21} \to +\infty$, which are in excellent agreement with the numerical results in the first row of Fig.~4.

For the mass-squared differences in the NO case, their evolution can be understood by using the RGEs in Eq.~(\ref{eq:dDeltaij}):
\begin{itemize}
\item For $\widetilde{\Delta}^{}_{21}$, the beta function is given by $|V^{}_{e2}|^2 - |V^{}_{e1}|^2$, which is initially negative but turns to be positive after crossing the first MSW resonance. This is why the ratio $\widetilde{\Delta}^{}_{21}/\Delta^{}_{21}$ gets its minimum at about $a/\Delta^{}_{21} = 1$. As $|V^{}_{e2}|^2$ increases rapidly to $1 - |U^{}_{e3}|^2$ afterwards and becomes stable until the second resonance is reached, $\widetilde{\Delta}^{}_{21}$ is linearly proportional to $a$ during this stable region. The ultimate value of $\widetilde{\Delta}^{}_{21}$ is fixed to $\Delta^{}_{31}$ for $a/\Delta^{}_{21} \to +\infty$.

\item For $\widetilde{\Delta}^{}_{31}$, the beta function is $|V^{}_{e3}|^2 - |V^{}_{e1}|^2$, which is negative as $|U^{}_{e3}|^2 \ll |U^{}_{e1}|^2$ at the beginning, so $\widetilde{\Delta}^{}_{31}$ decreases for the increasing $a$. But $|V^{}_{e1}|^2$ is reduced to zero quickly, while $|V^{}_{e3}|^2$ keeps almost unchanged, so the evolution of $\widetilde{\Delta}^{}_{31}$ is negligible. The situation changes when the second resonance is encountered and $|V^{}_{e3}|^2$ approaches one rapidly. Hence, $\widetilde{\Delta}^{}_{31}$ turns out to be linearly proportional to $a$ ultimately.
\end{itemize}
The results for antineutrinos and the IO case can be discussed in a similar way. As indicated in Fig.~\ref{fig:Jarlskog}, the Jarlskog invariant $\widetilde{\cal J}$ will be vanishing as $a/\Delta^{}_{21} \to +\infty$. This can be explained via the Naumov relation $\widetilde{\cal J} = {\cal J}\Delta^{}_{21} \Delta^{}_{31} \Delta^{}_{32}/(\widetilde{\Delta}^{}_{21} \widetilde{\Delta}^{}_{31} \widetilde{\Delta}^{}_{32})$, in which the denominator is approaching infinity. On the other hand, as $\widetilde{\theta}^{}_{13} \to \pi/2$ for $a/\Delta^{}_{21} \to +\infty$, the Jarlskog invariant is $\widetilde{\cal J} \propto \sin 2\widetilde{\theta}^{}_{13} \cos \widetilde{\theta}^{}_{13} \to 0$ in the standard parametrization.
\begin{figure}[!t]
\centering
\includegraphics[width=0.8\textwidth]{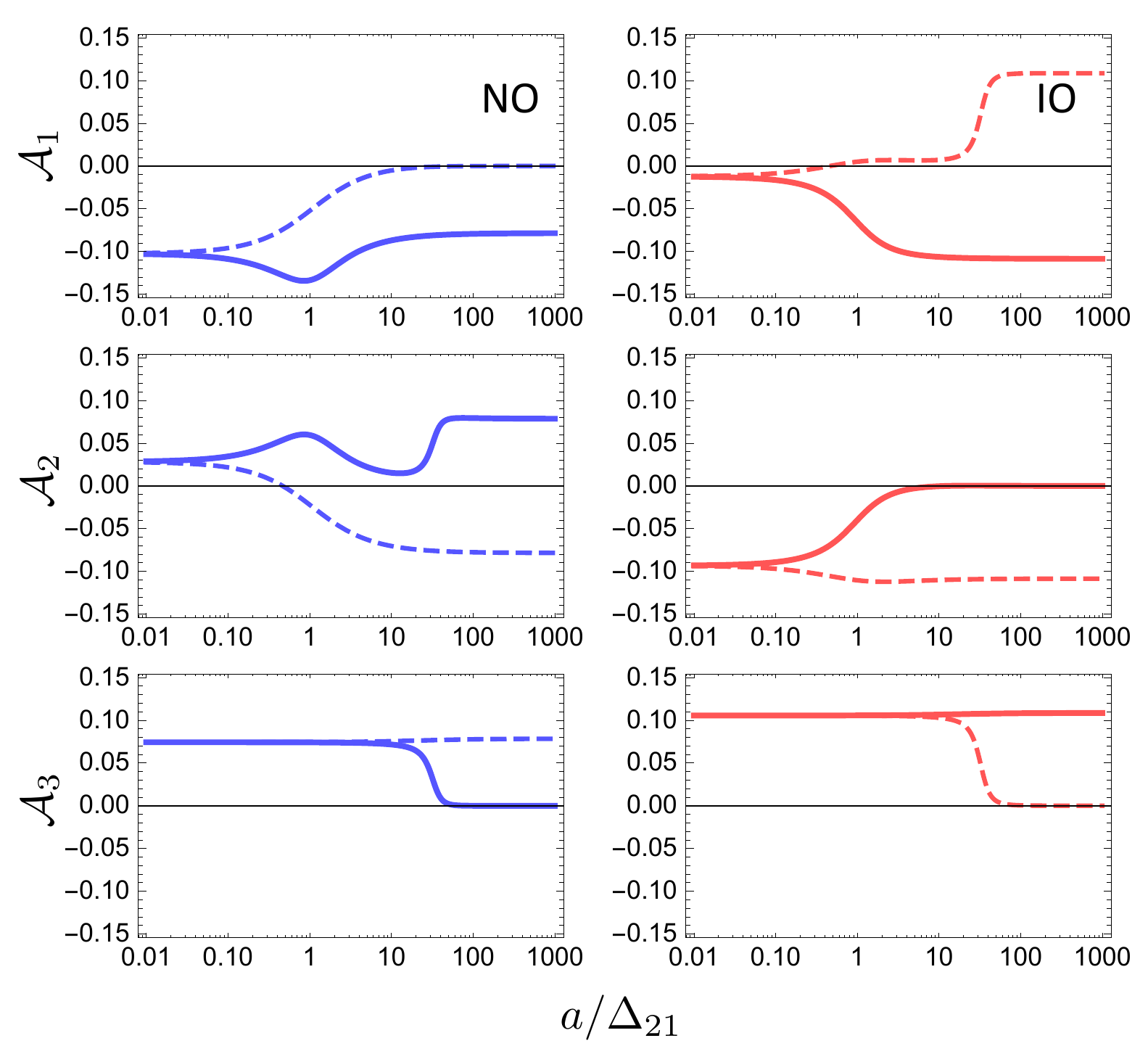}
\vspace{-0.3cm}
\caption{\label{fig:mu_tau} The evolution of the partial $\mu$-$\tau$ asymmetries $\mathcal{A}^{}_i \equiv |V^{}_{\mu i}|^2 - |V^{}_{\tau i}|^2$ for $i=1, 2 ,3$ with respect to $a/\Delta^{}_{21}$, where the same convention and input values as in Fig.~\ref{fig:Jarlskog} are taken. }
\end{figure}

Finally, one may wonder whether the RGEs in Eq.~(\ref{eq:anglephase})
can be analytically solved, so as
to express $\widetilde{\theta}^{}_{12}$, $\widetilde{\theta}^{}_{13}$, $\widetilde{\theta}^{}_{23}$
and $\widetilde{\delta}$ in terms of $\theta^{}_{12}$, $\theta^{}_{13}$, $\theta^{}_{23}$, $\delta$
and the relevant neutrino mass-squared differences. This will be a challenge if the matter
density is arbitrarily varying. Given a constant matter profile, however, the exact analytical relations between $\{\widetilde{\theta}^{}_{12}, \widetilde{\theta}^{}_{13}, \widetilde{\theta}^{}_{23}, \widetilde{\delta}\}$ and $\{\theta^{}_{12}, \theta^{}_{13}, \theta^{}_{23}, \delta\}$ have been established in Refs.~\cite{Zag, Xing1, Xing2} in a different approach. But those relations are so complicated that they are not very helpful for understanding
the behaviors of neutrino oscillations in matter. That is why some useful and more transparent analytical approximations have been made in the literature for long- and medium-baseline neutrino oscillation experiments (e.g., Ref.~\cite{Xing:2016ymg} for $E \lesssim 1$ GeV and Refs.~\cite{Cervera:2000kp, Freund:2001pn} for $E \gtrsim 0.5$ GeV).

\section{Some Further Discussions}

In this section we demonstrate that the RGEs derived in the previous sections can also be utilized to analyze the matter effects on several phenomenologically interesting observables. Let us begin with the partial $\mu$-$\tau$ asymmetries of $V$~\cite{Xing:2014zka},
\begin{eqnarray}\label{eq:Ai}
{\cal A}^{}_i  \equiv |V^{}_{\mu i}|^2 - |V^{}_{\tau i}|^2 \;
\end{eqnarray}
(for $i = 1, 2, 3$) and the off-diagonal asymmetries of $V$,
\begin{eqnarray}
\label{eq:AL}
{\cal A}^{}_{\rm L}  &\equiv& |V^{}_{e 2}|^2 - |V^{}_{\mu 1}|^2 = |V^{}_{\mu 3}|^2 - |V^{}_{\tau 2}|^2 = |V^{}_{\tau 1}|^2 - |V^{}_{e 3}|^2 \; , \nonumber\\
\label{eq:AR}
{\cal A}^{}_{\rm R}  &\equiv& |V^{}_{e 2}|^2 - |V^{}_{\mu 3}|^2 = |V^{}_{\mu 1}|^2 - |V^{}_{\tau 2}|^2 = |V^{}_{\tau 3}|^2 - |V^{}_{e 1}|^2 \; .
\end{eqnarray}
The phenomenological implications of the partial symmetry $|U^{}_{\mu 1}|^2 = |U^{}_{\tau 1}|^2$ or $|U^{}_{\mu 2}|^2 = |U^{}_{\tau 2}|^2$ for the leptonic CP-violating phase and mixing angles in vacuum have been investigated in Ref.~\cite{Xing:2014zka}, in which it has been shown that the leptonic CP-violating phase $\delta$ is correlated with three mixing angles if such a symmetry is imposed. In the standard parametrization, $|U^{}_{\mu 3}|^2 = |U^{}_{\tau 3}|^2$ leads to the maximal mixing angle $\theta^{}_{23} = \pi/4$. On the contrary, $|U^{}_{\mu 1}|^2 = |U^{}_{\tau 1}|^2$ or $|U^{}_{\mu 2}|^2 = |U^{}_{\tau 2}|^2$ allows for an appreciable deviation of $\theta^{}_{23}$ from $\pi/4$ and that of $\delta$ from $\pm \pi/2$, which are compatible with current neutrino oscillation data. Unlike the full $\mu$-$\tau$ symmetry $|V^{}_{\mu i}|^2 = |V^{}_{\tau i}|^2$, the partial symmetry is not preserved by matter effects, which can be seen from the following RGEs:
\begin{eqnarray}\label{eq:dAi}
\frac{{\rm d}}{{\rm d}a} {\cal A}^{}_i &=& \frac{{\rm d}}{{\rm d}a} |V^{}_{\mu i}|^2 - \frac{{\rm d}}{{\rm d}a} |V^{}_{\tau i}|^2 = \sum_{j \neq i} \left(|V^{}_{\mu i}|^2 |V^{}_{\mu j}|^2 - |V^{}_{\tau i}|^2 |V^{}_{\tau j}|^2\right) \widetilde{\Delta}^{-1}_{ji} \; ,
\end{eqnarray}
where Eq.~(\ref{eq:Vai2}) has been used. It is evident that if ${\cal A}^{}_i = 0$ (for $i = 1, 2, 3$) hold exactly in vacuum (namely, $|U^{}_{\mu i}|^2 = |U^{}_{\tau i}|^2$), they remain to be vanishing in matter. This point has also been emphasized in the previous section with the standard parametrization of $V$. However, if only the partial $\mu$-$\tau$ symmetry (say ${\cal A}^{}_1 = 0$ or $|U^{}_{\mu 1}|^2 = |U^{}_{\tau 1}|^2$) is valid in vacuum, then we have
\begin{eqnarray}\label{eq:dA1}
\frac{{\rm d}}{{\rm d}a} {\cal A}^{}_1 = 2|V^{}_{\mu 1}|^2 \left[ \left(|V^{}_{\mu 2}|^2 - |V^{}_{\tau 2}|^2\right) \widetilde{\Delta}^{-1}_{21} + \left(|V^{}_{\mu 3}|^2 - |V^{}_{\tau 3}|^2\right) \widetilde{\Delta}^{-1}_{31} \right] \; ,
\end{eqnarray}
which is in general nonzero for $|V^{}_{\mu 2}|^2 \neq |V^{}_{\tau 2}|^2$ and $|V^{}_{\mu 3}|^2 \neq |V^{}_{\tau 3}|^2$. Therefore, the predictions from $|U^{}_{\mu 1}|^2 = |U^{}_{\tau 1}|^2$ in vacuum are invalidated in matter. In a similar way, one can calculate the RGEs for the off-diagonal asymmetries,
\begin{eqnarray}\label{eq:dALR}
\frac{{\rm d}}{{\rm d}a} {\cal A}^{}_{\rm L} &=& 2 \left[{\rm Re}\left(V^{}_{\tau 1} V^{}_{e 2} V^*_{\tau 2} V^*_{e 1}\right) \widetilde{\Delta}^{-1}_{12} - |V^{}_{e 2}|^2 |V^{}_{e 3}|^2 \widetilde{\Delta}^{-1}_{23} + {\rm Re}\left(V^{}_{\mu 3} V^{}_{e 1} V^*_{\mu 1} V^*_{e 3}\right) \widetilde{\Delta}^{-1}_{31} \right]\; ,
\nonumber\\
\frac{{\rm d}}{{\rm d}a} {\cal A}^{}_{\rm R} &=& 2 \left[|V^{}_{e 1}|^2 |V^{}_{e 2}|^2 \widetilde{\Delta}^{-1}_{12}- {\rm Re} \left(V^{}_{\tau 2} V^{}_{e 3} V^*_{\tau 3} V^*_{e 2}\right) \widetilde{\Delta}^{-1}_{23} -  {\rm Re}\left(V^{}_{\mu 3} V^{}_{e 1} V^*_{\mu 1} V^*_{e 3}\right) \widetilde{\Delta}^{-1}_{31} \right] \; .
\end{eqnarray}
The latest neutrino oscillation data indicate that both ${\cal A}^{}_{\rm L}$ and ${\cal A}^{}_{\rm R}$ are nonzero for the flavor mixing matrix in vacuum.

\begin{figure}[!t]
\centering
\includegraphics[width=0.8\textwidth]{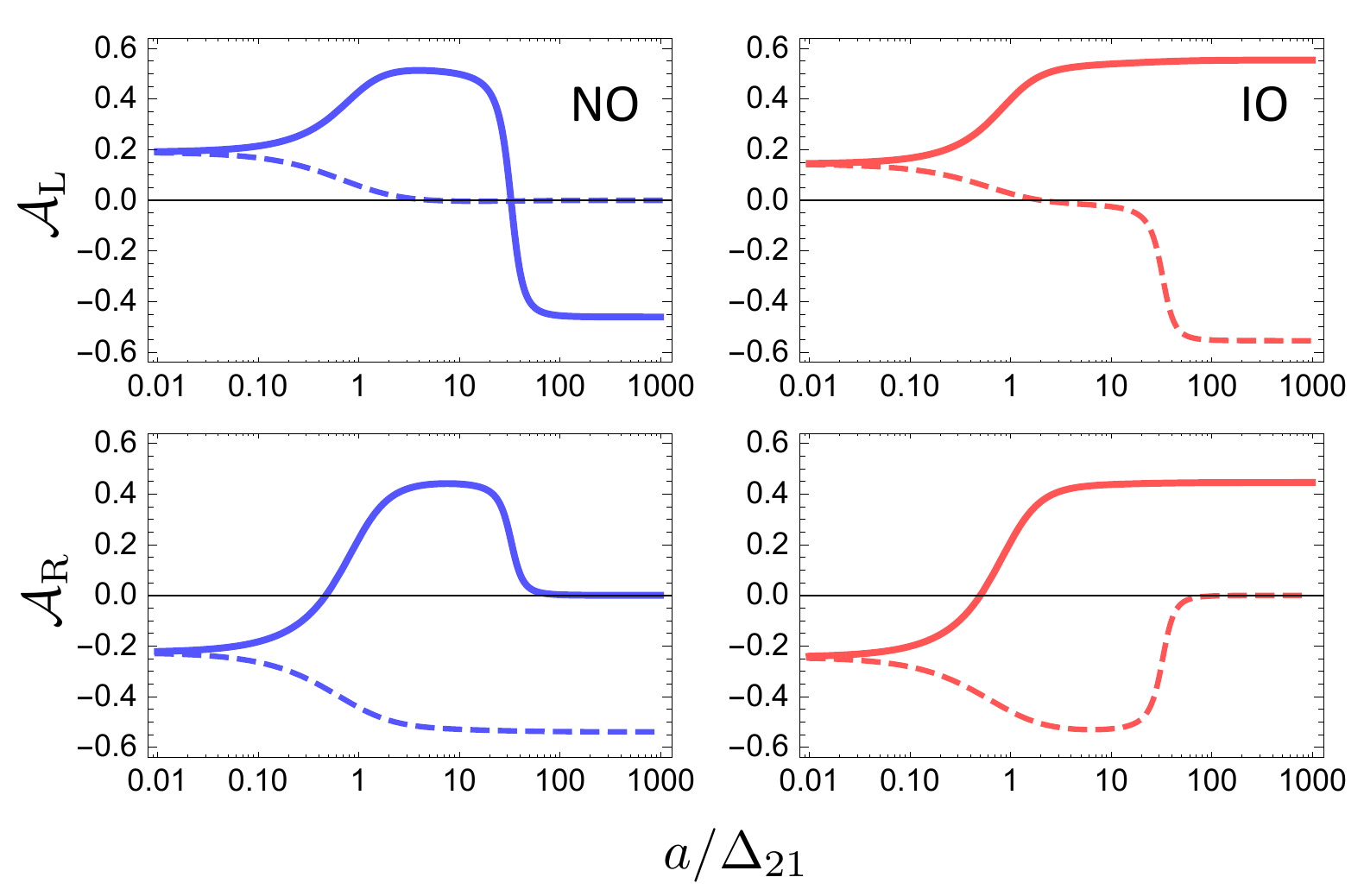}
\vspace{-0.3cm}
\caption{\label{fig:off_diagonal} The evolution of the off-diagonal asymmetries $\mathcal{A}^{}_{\text{L}}$ and $\mathcal{A}^{}_{\text{R}}$ with respect to $a/\Delta^{}_{21}$, where the same convention and input values as in Fig.~\ref{fig:Jarlskog} are taken.}
\end{figure}

We present the running behaviors of the partial $\mu$-$\tau$ asymmetries and the off-diagonal asymmetries in Figs.~\ref{fig:mu_tau} and \ref{fig:off_diagonal}, respectively. Assuming the initial values of neutrino mixing parameters in vacuum to be the best-fit numbers, one can find that the asymmetries $|{\cal A}^{}_i| \lesssim 0.1$, which are indeed modified by the matter effects, but only slightly. On the other hand, however, the off-diagonal asymmetries can be significantly enhanced or suppressed during the evolution with respect to $a/\Delta^{}_{21}$. It is straightforward to explain the primary features of the evolution of these asymmetries by using the numerical results in Figs.~\ref{fig:mixing_NO} and \ref{fig:mixing_IO}.

Next, we focus on the sides of six leptonic unitarity triangles of $V$, which are defined by the orthogonality conditions in the complex plane \cite{FX}:
\begin{eqnarray}\label{eq:UT1}
&& \triangle^{}_e: V^{}_{\mu 1} V^*_{\tau 1} + V^{}_{\mu 2} V^*_{\tau 2} + V^{}_{\mu 3} V^*_{\tau 3} = 0 \; , \nonumber \\
&& \triangle^{}_\mu: V^{}_{\tau 1} V^*_{e 1} + V^{}_{\tau 2} V^*_{e 2} + V^{}_{\tau 3} V^*_{e 3} = 0 \; , \nonumber \\
&& \triangle^{}_\tau: V^{}_{e 1} V^*_{\mu 1} + V^{}_{e 2} V^*_{\mu 2} + V^{}_{e 3} V^*_{\mu 3} = 0 \; ;
\end{eqnarray}
and
\begin{eqnarray}\label{eq:UT2}
&& \triangle^{}_1: V^{}_{e 2} V^*_{e 3} + V^{}_{\mu 2} V^*_{\mu 3} + V^{}_{\tau 2} V^*_{\tau 3} = 0 \; , \nonumber \\
&& \triangle^{}_2: V^{}_{e 3} V^*_{e 1} + V^{}_{\mu 3} V^*_{\mu 1} + V^{}_{\tau 3} V^*_{\tau 1} = 0 \; , \nonumber \\
&& \triangle^{}_3: V^{}_{e 1} V^*_{e 2} + V^{}_{\mu 1} V^*_{\mu 2} + V^{}_{\tau 1} V^*_{\tau 2} = 0 \; .
\end{eqnarray}
Taking the unitarity triangle $\triangle^{}_\tau$ for example, one may figure out
\begin{eqnarray}\label{eq:UTtau}
\frac{{\rm d}}{{\rm d}a} \left(V^{}_{e 1} V^*_{\mu 1}\right) = V^{}_{e 1} V^*_{\mu 1}\left(|V^{}_{e2}|^2 \widetilde{\Delta}^{-1}_{12} - |V^{}_{e3}|^2 \widetilde{\Delta}^{-1}_{31}\right) + |V^{}_{e1}|^2 \left(V^{}_{e2}V^*_{\mu 2}\widetilde{\Delta}^{-1}_{12} - V^{}_{e3}V^*_{\mu 3}\widetilde{\Delta}^{-1}_{31}\right) \; , \nonumber \\
\frac{{\rm d}}{{\rm d}a} \left(V^{}_{e 2} V^*_{\mu 2}\right) = V^{}_{e 2} V^*_{\mu 2}\left(|V^{}_{e3}|^2 \widetilde{\Delta}^{-1}_{23} - |V^{}_{e1}|^2 \widetilde{\Delta}^{-1}_{12}\right) + |V^{}_{e2}|^2 \left(V^{}_{e3}V^*_{\mu 3}\widetilde{\Delta}^{-1}_{23} - V^{}_{e1}V^*_{\mu 1}\widetilde{\Delta}^{-1}_{12}\right) \; , \nonumber \\
\frac{{\rm d}}{{\rm d}a} \left(V^{}_{e 3} V^*_{\mu 3}\right) = V^{}_{e 3} V^*_{\mu 3}\left(|V^{}_{e1}|^2 \widetilde{\Delta}^{-1}_{31} - |V^{}_{e2}|^2 \widetilde{\Delta}^{-1}_{23}\right) + |V^{}_{e3}|^2 \left(V^{}_{e1}V^*_{\mu 1}\widetilde{\Delta}^{-1}_{31} - V^{}_{e2}V^*_{\mu 2}\widetilde{\Delta}^{-1}_{23}\right) \; ,
\end{eqnarray}
where Eq.~(\ref{eq:elements}) has been utilized to compute the derivatives of the matrix element and its complex conjugate. From Eq.~(\ref{eq:UTtau}), we can observe how the three sides of $\triangle^{}_\tau$ are changing with the matter term. Since the evolution of all the six leptonic unitarity triangles has been systematically studied in Refs.~\cite{Zhang:2004hf, Xing:2005gk}, we do not elaborate on this issue here.

\begin{figure}[!t]
\centering
\includegraphics[width=0.6\textwidth]{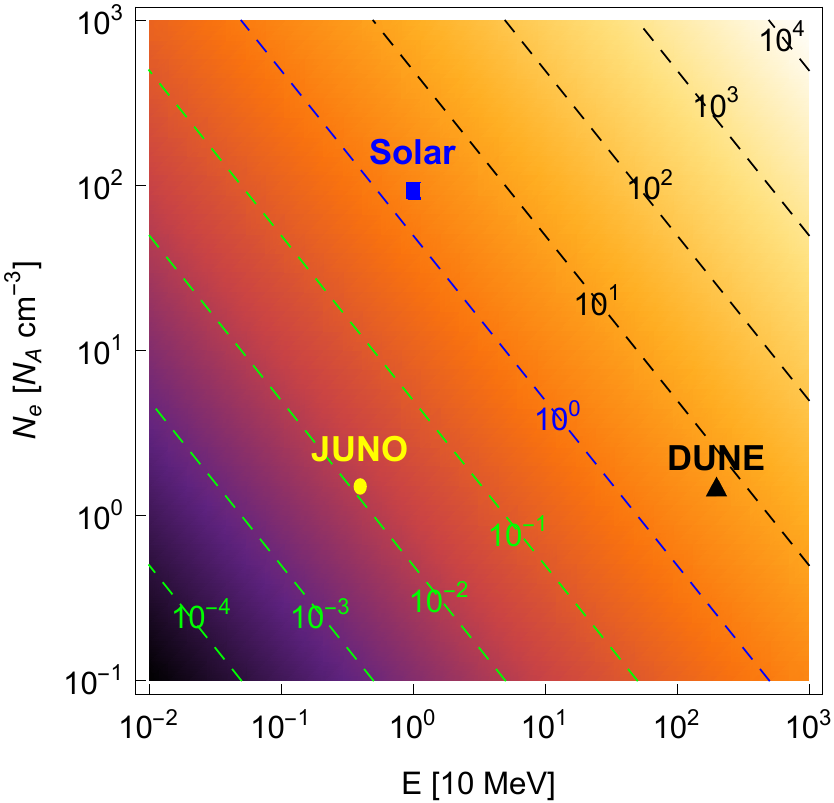}
\vspace{-0.2cm}
\caption{\label{fig:density} The contours of $a/\Delta^{}_{21} = 0.02~\cdot [N^{}_e/(N^{}_{\rm A}~{\rm cm}^{-3})] \cdot [E/(10~{\rm MeV})]$ in the plane of neutrino energy $E$ and the net electron number density $N^{}_e$, where the yellow disk stands for JUNO at $(E, N^{}_e) = (4~{\rm MeV}, 1.5 N^{}_{\rm A}/{\rm cm}^3)$, the blue square for solar neutrinos at $(10~{\rm MeV}, 10^2 N^{}_{\rm A}/{\rm cm}^3)$, and the black triangle for DUNE at $(2~{\rm GeV}, 1.5 N^{}_{\rm A}/{\rm cm}^3)$.}
\end{figure}

Last but not least, we give some remarks on the parameter $a/\Delta^{}_{21}$, which has been chosen as an arbitrary dimensionless scale-like variable. Based on the definition $a \equiv 2\sqrt{2} \ G^{}_{\rm F} N^{}_e E$, it is convenient to rewrite $a/\Delta^{}_{21}$ as follows:
\begin{eqnarray}\label{eq:aexp}
\frac{a}{\Delta^{}_{21}} = 0.02~\left(\frac{N^{}_e}{N^{}_{\rm A}~{\rm cm}^{-3}}\right) \cdot \left(\frac{E}{10~{\rm MeV}}\right) \; ,
\end{eqnarray}
where $N^{}_{\rm A} = 6.022\times 10^{23}$ is the Avogadro constant, and the electron number density $N^{}_e$ is related to the matter density $\rho$ through $N^{}_e = N^{}_{\rm A}~{\rm cm}^{-3}~Y^{}_e~[\rho/(1~{\rm g}~{\rm cm}^{-3})]$. In Fig.~\ref{fig:density} the contours of $a/\Delta^{}_{21}$ have been shown in the plane of $(E, N^{}_e)$, and three typical neutrino oscillation experiments have been indicated on the plot for the purpose of illustration: JUNO reactor antineutrinos at $(4~{\rm MeV}, 1.5 N^{}_{\rm A}/{\rm cm}^3)$ \cite{JUNO}, solar neutrinos at $(10~{\rm MeV}, 10^2 N^{}_{\rm A}/{\rm cm}^3)$ \cite{Patrignani:2016xqp} and DUNE with accelerator neutrinos at $(2~{\rm GeV}, 1.5 N^{}_{\rm A}/{\rm cm}^3)$ \cite{DUNE}. For the reactor- and accelerator-based experiments, the matter density is usually taken to be $\rho = 3~{\rm g}~{\rm cm}^{-3}$, i.e., the average density of the Earth crust or mantle. Thus, the evolution with respect to $a$ can be realized by changing the neutrino beam energy or the matter density.

As a potentially interesting application of the RGE approach developed above, one may first express the neutrino oscillation probabilities relevant for those realistic experiments in terms of the effective mixing parameters and then extract their values directly from the corresponding experimental data. The exact RGEs of those effective mixing parameters can subsequently be implemented to run the measured values to a common scale of $a/\Delta^{}_{21}$. In particular, the fundamental oscillation parameters (i.e., two neutrino mass-squared differences and four flavor mixing parameters) can be extrapolated from their matter-corrected counterparts in the vacuum limit of $a/\Delta^{}_{21} \to 0$. It is still unclear whether this procedure will work better than the usual treatment of matter effects in the present neutrino oscillation experiments with reasonable analytical approximations, but its principle is definitely on solid ground because the language of RGEs itself is completely model-independent.

\section{Concluding Remarks}

It is well known that the RGE approach has been serving as a powerful tool in a number of
aspects of theoretical physics to systematically describe the changes of
a physical system as viewed at different distances or energy scales, and its
success in quantum field theory is especially marvelous.
In the present work we have applied this language
to the description of neutrino masses and flavor mixing parameters in a medium,
which evolve with the arbitrary scale-like matter parameter
$a \equiv 2\sqrt{2} \ G^{}_{\rm F} N^{}_e E$, and highlighted a striking
possibility that the genuine neutrino flavor quantities in vacuum can be
extrapolated from their matter-corrected counterparts to be measured in
some realistic neutrino oscillation experiments.

To be explicit, we have clearly demonstrated that the dependence of the effective flavor mixing parameters $V^{}_{\alpha i}$ and $\widetilde{m}^2_i$ on the matter parameter $a$ can perfectly be described by a complete set of differential equations, which are just referred to as the RGEs of those quantities. The point is that the introduction of effective neutrino mass-squared
differences and flavor mixing parameters guarantees the form invariance or self-similarity
of neutrino oscillation probabilities in vacuum and in matter, and hence the RGE-like approach for describing neutrino oscillations in matter works well. In addition to the RGEs for $\widetilde{m}^{}_i$ and $|V^{}_{\alpha i}|^2$~\cite{Chiu:2010da, Chiu:2017ckv}, we have also derived the RGEs of three flavor mixing angles and one CP-violating phase in the standard parametrization of $V$, and numerically illustrated some salient features of their evolution with respect to the matter parameter $a$. The RGEs of $\widetilde{\cal J}$ and some other interesting quantities, such as the partial $\mu$-$\tau$ asymmetries, the off-diagonal asymmetries and the sides of unitarity triangles of $V$, have been derived and discussed as a by-product of this work.
The Naumov and Toshev relations are reformulated too.

In the long run, the RGE-like approach that we have developed may hopefully provide a generic framework for the systematic study of neutrino masses and flavor mixing parameters in any possible
matter environments. Although such a tool might be ``scientifically indistinguishable" from the conventional methods of dealing with matter effects on neutrino oscillations, ``they are not psychologically identical" in making the underlying physics more transparent \cite{Feynman}. In particular, tracing an analogy between the evolution of neutrino masses and flavor mixing parameters in matter and their evolution with the energy scale is theoretically interesting. We therefore expect that our work can find some useful applications in neutrino phenomenology.

\vspace{0.5cm}

\textsl{This work was supported in part by
the National Natural Science Foundation of China under grant No. 11775231 (XZZ) and grant No. 11775232 (SZ), by the National Youth Thousand Talents Program (SZ), by the CAS Center for Excellence in Particle Physics (SZ), and by the European Research Council under ERC Grant NuMass (FP7-IDEAS-ERC ERC-CG 617143).}

\end{document}